\documentclass[conference]{IEEEtran}
\usepackage{array}

\newcolumntype{R}[1]{>{\raggedleft\let\newline\\\arraybackslash\hspace{0pt}}m{#1}}
\newcolumntype{L}[1]{>{\raggedright\let\newline\\\arraybackslash\hspace{0pt}}m{#1}}

\ifCLASSINFOpdf
  \usepackage[pdftex]{graphicx}
\else
\fi
\ifCLASSOPTIONcompsoc
  \usepackage[caption=false,font=normalsize,labelfont=sf,textfont=sf]{subfig}
\else
  \usepackage[caption=false,font=footnotesize]{subfig}
\fi
\begin{document}
%
\title{Co-location Epidemic Tracking on London Public Transports Using Low Power Mobile Magnetometer}

\author{\IEEEauthorblockN{Khuong An Nguyen}
\IEEEauthorblockA{Department of Computer Science\\
Royal Holloway, University of London\\
Surrey, United Kingdom\\
Email: Khuong.Nguyen.2007@live.rhul.ac.uk}
\and
\IEEEauthorblockN{Chris Watkins}
\IEEEauthorblockA{Department of Computer Science\\
	Royal Holloway, University of London\\
	Surrey, United Kingdom\\
	Email: C.J.Watkins@rhul.ac.uk}
\and
\IEEEauthorblockN{Zhiyuan Luo}
\IEEEauthorblockA{Department of Computer Science\\
	Royal Holloway, University of London\\
	Surrey, United Kingdom\\
	Email: Zhiyuan.Luo@rhul.ac.uk}
}


%


\maketitle

\begin{abstract}
The public transports provide an ideal means to enable contagious diseases transmission. This paper introduces a novel idea to detect co-location of people in such environment using just the ubiquitous geomagnetic field sensor on the smart phone. Essentially, given that all passengers must share the same journey between at least two consecutive stations, we have a long window to match the user trajectory. Our idea was assessed over a painstakingly survey of over 150 kilometres of travelling distance, covering different parts of London, using the overground trains, the underground tubes and the buses.
\end{abstract}


%
\IEEEpeerreviewmaketitle

\section{Introduction}
In 2015, it was reported that over 3 millions people relied on public transports in London every day, with an average of 45 minutes on board per person\footnote{https://www.gov.uk/government/statistics/transport-statistics-great-britain-2015 - last accessed in Feb/2017}. Such condition is ideal for infectious diseases to spread. For instance, an ill person's openly sneeze or cough may easily spread to other fellow passengers on a poorly ventilated underground tube in a long journey. Thus, co-location detection of people in such highly infectious environment is critical to control or predict the disease spreading rate in an event of epidemic.

For the past decade, the emerging of mobile devices provides a unique opportunity to tackle this challenge, since most people carry a smart phone with them when they are out and about. More importantly, every mobile device is equipped with multiple sensors that are capable of passively scanning the surroundings. However, little work was done within the health research community to make use of these sensors' reading. In this paper, we propose the use of the geomagnetic field sensor (magnetometer) to detect co-location of people on the public transports. We assume that, when two mobile devices observe similar time-stamped sensor' readings, they should be nearby, which in turns, guarantees that their respective owners should also be close by. Critically, since every passenger must share the same journey between at least two consecutive stations, which may last up to 10 minutes on the trains or buses, we have a window of opportunity to assess co-location of people.

The foremost advantage of our approach is that, at the time of writing, Google consider magnetometer to be a low power basic sensor, and thus, allowing it to be always-on and can be inquired without any permission, even in flight-safe mode. This is important for any passive epidemic tracking app to run seamlessly without the hassle of asking for the user permission (e.g. Since Android 6.0, Google demand any app that uses WiFi or Bluetooth to ask for real-time permission to access the user location).

Overall, the paper identifies the following contributions:
\begin{itemize}
	\item We propose the use of magnetism to detect co-location of people. No wireless signals (e.g. WiFi, Bluetooth, GPS, Cellular) are needed.
	\item We detail our algorithm to robustly detect same-carriage co-localisation.
	\item We assess our system in large scale real-world settings which cover 150 kilometres of travelling distance in different parts of London, on all types of public transports (i.e. the overground trains, the underground tubes, and the buses).
\end{itemize}

The remaining of the paper is organised into six sections. Section II tells the story behind our ideas of using magnetism. So that, Section III can build on to explain our concept of magnetic based co-location, emphasising on the challenges facing such approach. Then, Section IV details the experiments including the test environments and the empirical results. Section V overviews other related work. Lastly, Section VI summaries our work and outlines further research.

\section{Magnetism based co-location of people}
This section justifies the selection of magnetism for this paper and compares it to other wireless based competitors.

\subsection{An inspiration and opportunity of using magnetism for localisation}
\begin{figure}[!t]
	\centering
	
	\subfloat[High level of magnetism distortion during a 16 minute train journey]{\includegraphics[width=3.0in]{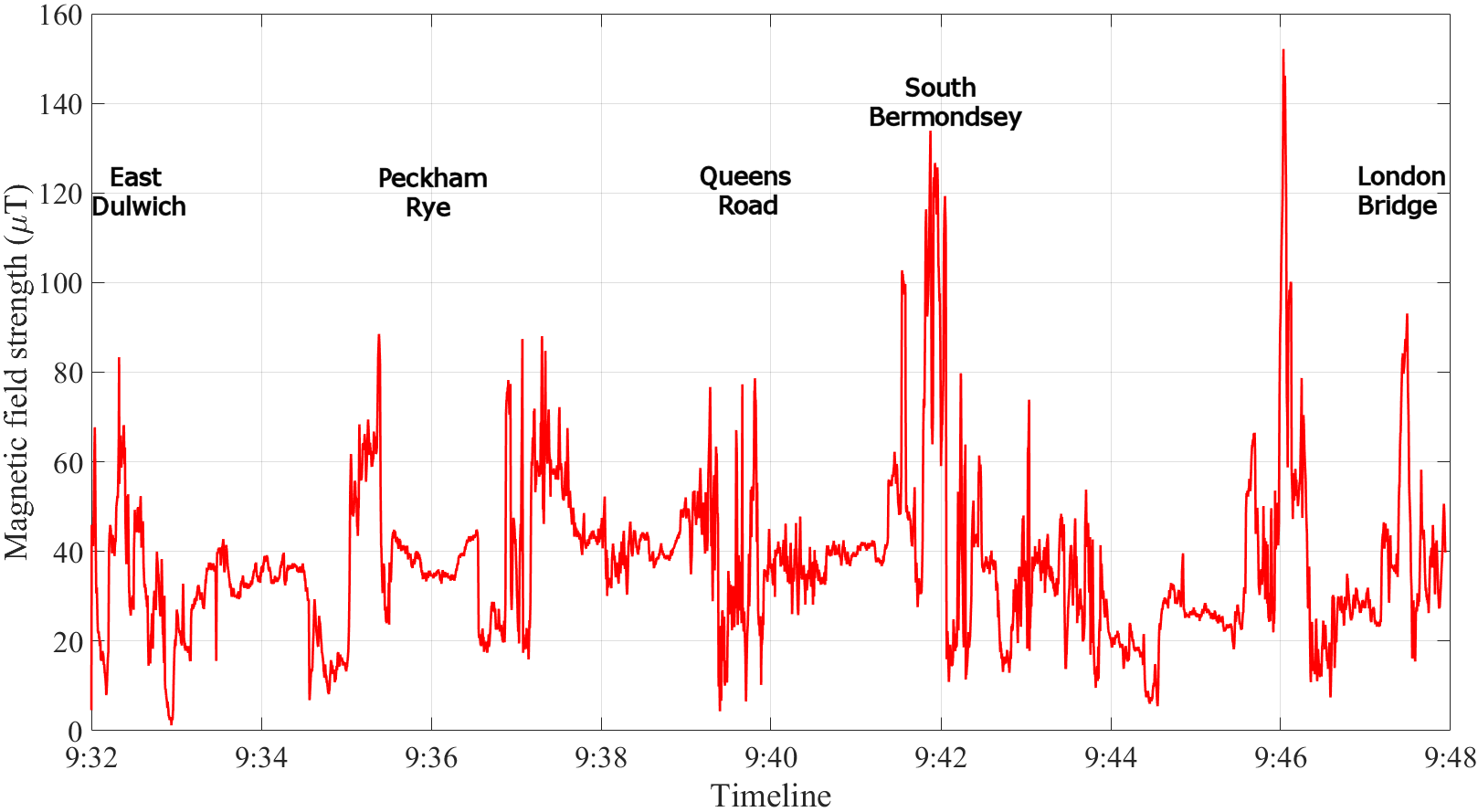}
		\label{dulwichbridgelabel}}
	\hfil
	\subfloat[The heatmap visualisation of the trip]{\includegraphics[width=2.0in]{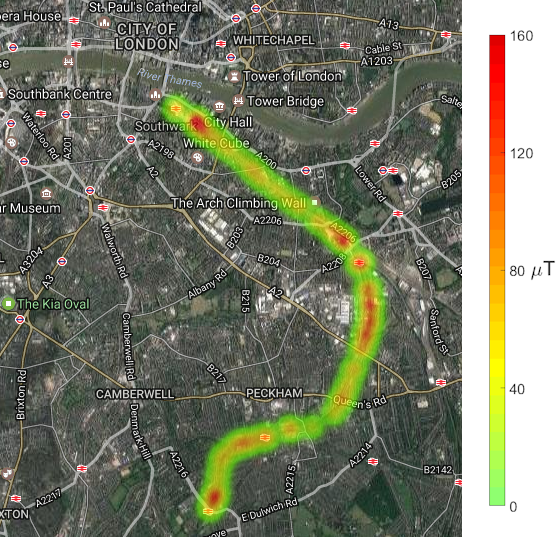}
		\label{heatmapdulwichbridge}}
	\hfil
	
	\caption{An inspiration for using magnetism for co-location detection on public transports. The magnetism observed on-board of an overground train from South-East to Central London, passing through 5 stations is heavily distorted.}
	\label{magnetismdulwichbridge}
\end{figure}

It is well-known that animals rely on the Earth's magnetic field to perform route-finding in nature (e.g. the birds know where to head North in migratory season). Regrettably, such technique cannot be applied indoors or undergrounds, because the natural magnetic field generated by the Earth's core is heavily distorted by the metal bars, steel rebars, ferrous tubes and reinforced concrete which are commonly found within the building structure. Additionally, an electric current that moves in metal wires (e.g. power lines) will also alter the nearby magnetic field. However, this challenge provides a `unique' opportunity for the purpose of co-location detection. That is, the magnetic field is not uniformly perturbed, so that, different areas experience different magnetism anomalies (see Figure~\ref{magnetismdulwichbridge}).

Nevertheless, the ultimate research question is: \textbf{To what extent can magnetism be used to differentiate two separate positions?} For the purpose of epidemic tracking, we are looking at city-level operation, and it is unavoidable that several locations may exhibit the same magnetic signature. There are four reasons that inspire us to venture towards this approach. 
\begin{enumerate}
	\item We are only interested in co-localisation, that is, the exact moment two persons are close by. As such, a time stamp constraint will get rid of most similar samples collected at different times. 
	\item We focus our attention on the public transports, which guarantee that all passengers must follow the same trajectory for at least two consecutive stations. This window supplies a long sequence of samples which allows us to further differentiate non-co-located users. 
	\item Modern public transports are electric-based (e.g. those used in London) which greatly alter the on-board magnetic field area. Additionally, other trains that run on adjacent tracks may temporarily distort the magnetism of the neighbourhood trains.
	\item The ferrous structure from nearby buildings may have a unique magnetic signature that all passengers on the same train must observe, albeit with different time delays (i.e. the passenger at the front of the train will `see' the building a few seconds earlier than the one at the back).
\end{enumerate}

\subsection{Pros and cons of using magnetism}
For the purpose of co-location that leverages the smart phone's sensors, the magnetic field strength is not the only viable option. Other popular wireless signals such as Bluetooth, WiFi, Cellular, GPS have their own pros and cons (see Table~\ref{comparison}). 
\begin{table*}
	\caption{Comparison of smart phone's sensors for co-location purpose.}
	\centering 
	\begin{tabular}{| c | c | c | c | c | c |} 
		\hline\hline 
		& \textbf{Magnetometer} & \textbf{WiFi} & \textbf{Bluetooth} & \textbf{Cellular} & \textbf{GPS} \\ [0.5ex] 
		\hline 
		Coverage & Ubiquitous & Mostly indoors \& City centrals & Indoors & Urban areas & Outdoors \\
		Ease of access & No permission & Need user permission & Need user permission & Need user permission & Need user permission \\
		Power consumption & Low & High & Low & Average & Very high \\
		Sampling rate & 49.65 Hz & 2 Hz & 1 Hz & 0.1 Hz & 1 Hz  \\
		Spatial uniqueness & Changing & Changing & Changing & Changing & High \\
		Temporal variation & Low & High & High & High & Low \\
		\hline 
	\end{tabular}
	\label{comparison}
\end{table*}

Coverage-wise, the magnetic field is available anywhere on Earth, whereas, GPS, WiFi, Bluetooth and Cellular wireless signal depend on the distance to nearby stations or satellites. In terms of power level, five hours of magnetometer's continuous inquiry plus writing the results to a file consumes as little as 7\% of battery, according to the in-built Android power measure, compared to over 45\% of that using GPS, and 30\% using WiFi. As a matter of fact, Android even allows the magnetometer to function normally in both `Flight safe' mode and `Power saving' mode, where most other sensors are suppressed or turned off completely. Additionally, the magnetometer achieves a fine-grained sampling rate at about 49.65 Hz with both of our test phones (about 50 samples per second), compared to just 3 samples per second with Bluetooth or about 1.5 samples with WiFi. It is worth noting that since Android offers 3 levels of magnetometer sampling - 4.96 Hz, 14.89 Hz and 49.65 Hz, we opted for the fastest one. This is essential for the underground tube test scenario, where the average speed of the tube is 60 kilometres per hour. Lastly, the ease of access is probably the most overlooked strength of the magnetometer, for which no permission whatsoever is required from either the user or the app to inquire the sensor's readings, at the time of writing.
\begin{figure}[!t]
	\centering
	\includegraphics[width=1.5in]{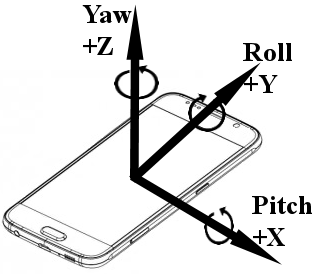}
	\caption{The three axes measured by the magnetometer.}
	\label{phone3d}
\end{figure}

However, despite these apparent benefits, the magnetic field strength is not spatially unique, because it contributes just 3 measures at each position, corresponding to the strength along each of the 3 axes (see Figure~\ref{phone3d}). In contrast, WiFi or Cellular based solutions have a much richer positioning representation, since they obtain references from several nearby stations. More problematically, the 3D orientation of the phone varies the above 3 measures. As such, the 3 measures must be reduced into one scalar magnitude, which practically means we only have 1 magnetic field based measure for every position.

\section{Analysing the sensor's footprints for co-location detection}
\begin{figure}[!t]
\centering
\includegraphics[width=1.2in]{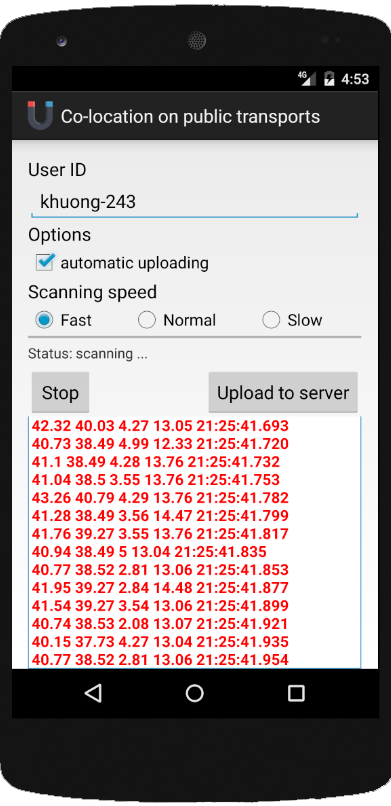}
\caption{The Android app used to collect the magnetic field strength.}
\label{app}
\end{figure}

Now we are in a good position to explain our co-location detection idea. At the beginning, the user installs an Android app on their device (see Figure~\ref{app}). The app's mission is to silently collect the magnetic field strength in the background. Each magnetic reading is accompanied by a time stamp and an activity recognition parameter, which will be discussed shortly. In an event of epidemic, the user submits his personal sensor data to a central server, which also manages other users' data. The process of co-location detection will be performed by comparing each pair of user data as follows. 

Without loss of generality, let us assume the first user - Alice submits her data in the form of $(\vec{p_1}, \dots, \vec{p_N})$, where $\vec{p_i}=(m_i, a_i, t_i)$ is the representing vector of position $i^{th}$ on Alice's journey comprising of $N$ positions. $m_i$ is the scalar magnitude reported by the magnetometer and $a_i$ is the recognised activity (to be discussed below) at time $t_i$ $(1 \leq i \leq N)$. The second user - Bob's trajectory is in a similar format of $(\vec{p'_1}, \dots, \vec{p'_M})$. Our objective is to verify whether Alice and Bob were co-located, and if so, when did that happen.

\textbf{Step 1: Smoothing the data}

\begin{figure}[!t]
	\centering
	\includegraphics[width=3.5in]{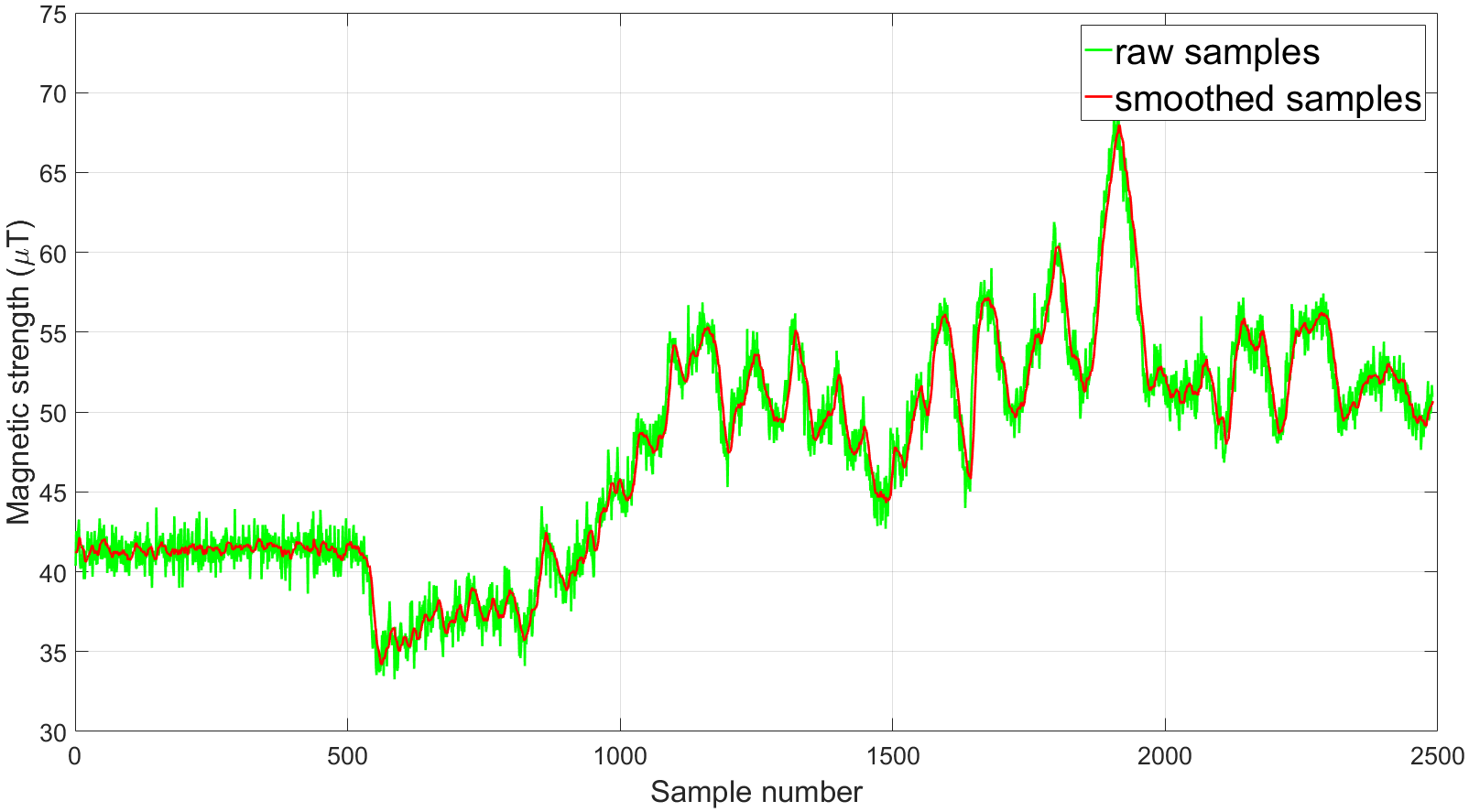}
	\caption{A part of the magnetic samples with/without the moving average filter. The filter reduces the overall electric noises from the magnetometer.}
	\label{movingfilter}
\end{figure}

We applied a linear moving average filter on the magnetometer outputs to smooth out the short-term electric noises from the sensor and to expose the true magnetic changes generated from the vehicle and the environment (see Figure~\ref{movingfilter}). An empirical window size filter of 10 was applied, since we can acquire up to 50 samples per second. Without loss of generality, given a sequence of magnetic readings $(m_1, \dots, m_N) (11 \leq i \leq N)$, with $N$ is the length of the sequence, each magnetic sample is smoothed out as follows.
\begin{equation}
m_i = \frac{\sum\limits_{j=1}^{10}{m_{i-j}}}{10}
\end{equation}

\begin{figure*}[!t]
	\centering
	
	\subfloat[Euclidean alignment]{\includegraphics[width=2.0in]{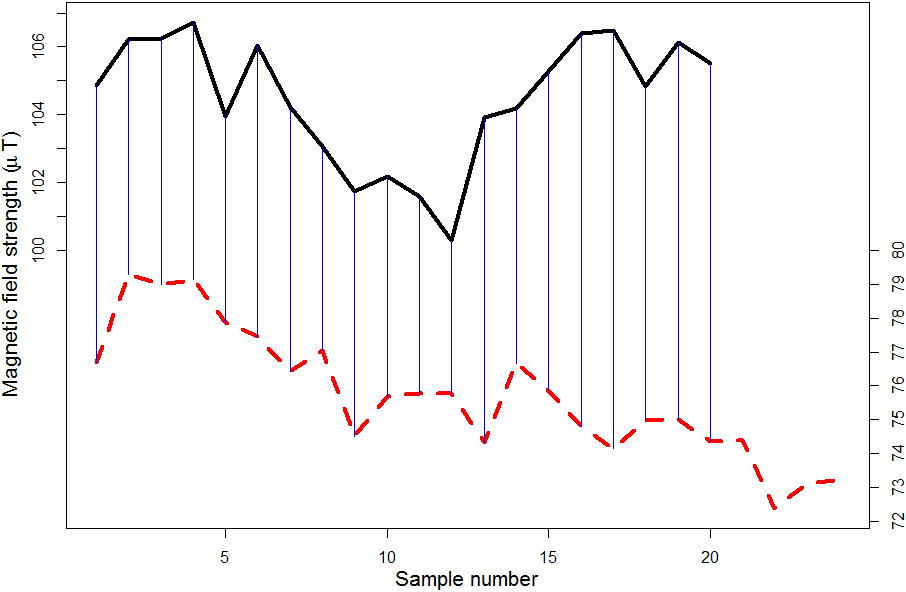}
		\label{euclideanalignment}}
	
	\subfloat[DTW alignment]{\includegraphics[width=2.0in]{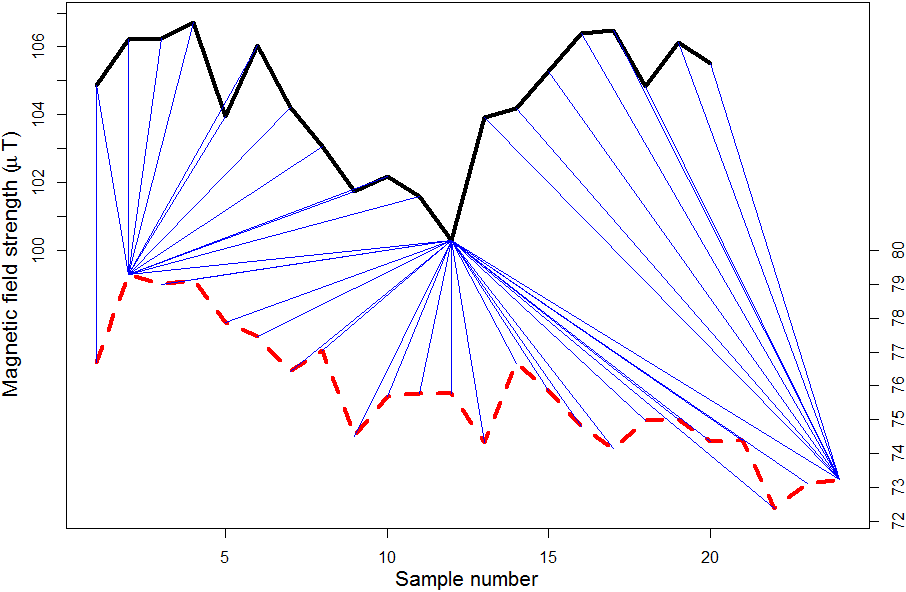}
		\label{dtwalignment}}
	\hfil
	\subfloat[DTW warping path]{\includegraphics[width=1.5in]{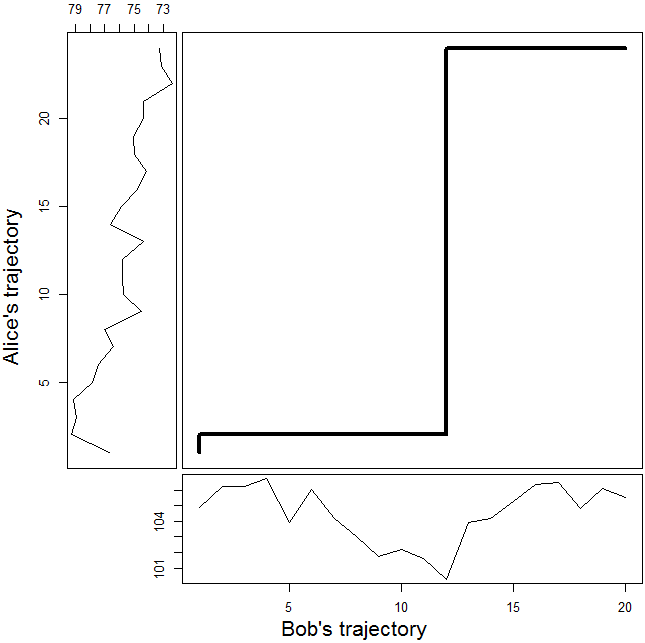}
		\label{dtwwarping}}
	
	\subfloat[Derivative DTW alignment]{\includegraphics[width=2.0in]{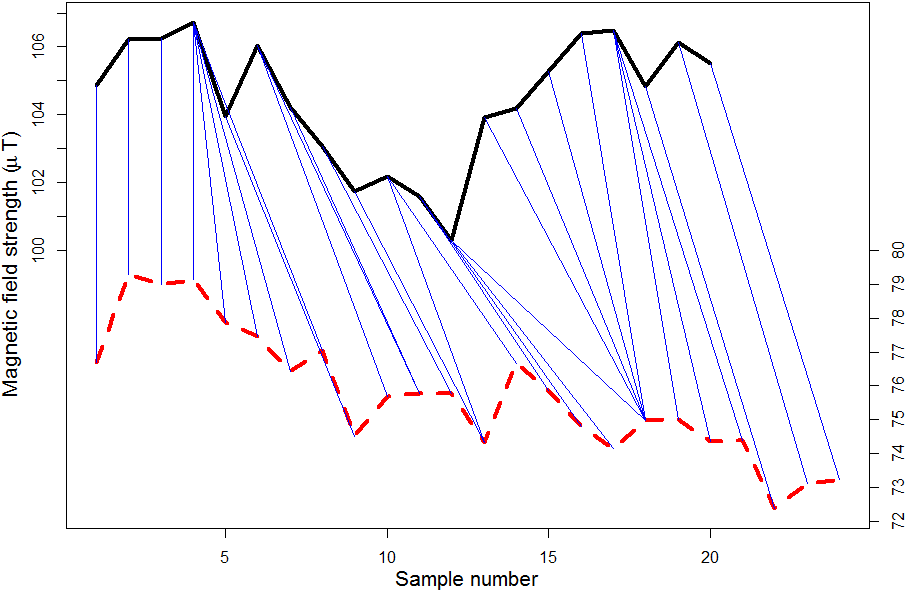}
		\label{ddtwalignment}}
	\hfil
	\subfloat[Derivative DTW warping path]{\includegraphics[width=1.5in]{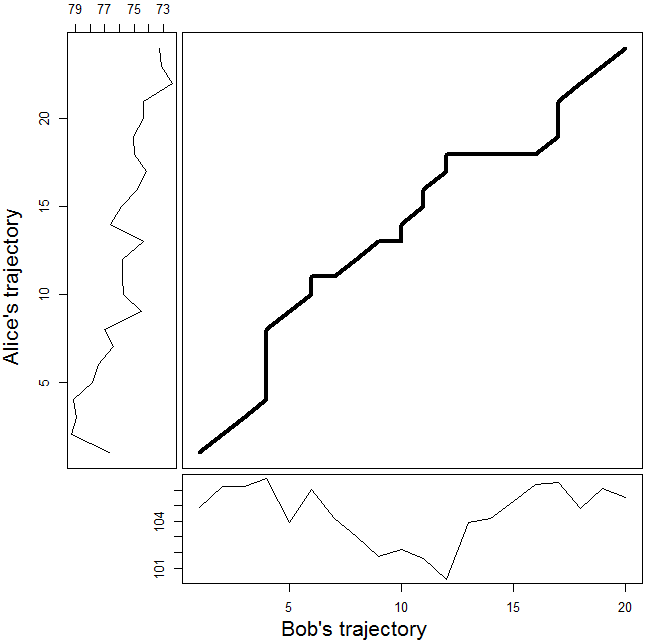}
		\label{ddtwwarping}}
	\hfil
	
	\caption{The justification for using Derivative DTW. Euclidean distance based matching fails to align trajectories of different lengths, while standard DTW over-warps the X-axis to explain the variability of the Y-axis.}
	\label{euclideanvsdtw}
\end{figure*}

\textbf{Step 2: Filtering the public transport related sequences}

The user's sensor data reflects his continuous activities through out the whole recorded period. However, we are only interested in parts of the data where the public transports were used. Hence, we employ the Activity Recognition API provided by Android to extract those\footnote{https://developers.google.com/android/reference/com/google/android/gms/\\location/ActivityRecognitionApi - last accessed in Feb/2017.}. This process runs in real-time along side with the data collection. The crux of this algorithm is that it uses a Bayesian classifier over the accelerometer readings to decide the likelihood of the current activity. Eight different activities are currently supported (i.e. Walking, Running, Still, On Foot, On Bicycle, In Vehicle, Tilting and Unknown). For our purpose, we are only interested in two main activities, that are, `In Vehicle' and `On Foot'. A magnetic sequence will be extracted if it begins with an `On Foot' event, following by an `In Vehicle' event, which signals that the user is entering the train or bus, and ends with another `On Foot' event, which signals that the user is leaving the vehicle. At the end of this step, each user's data is split into multiple trajectories, where each of them represents a separate trip on a public transport.

\textbf{Step 3: Finding the pair of matched trajectories}

Each of Alice's trajectory will be compared to all of Bob's trajectories to determine if they were co-located. The reverse process is unnecessary since the relationship is both-sided. We employed Derivative Dynamic Time Warping (DDTW)~\cite{keogh2001derivative} to match two magnetic trajectories for four reasons. 

Firstly, it stretches the shorter trajectory to match the longer one, which is essential for our purpose because of the sensor's delay to always guarantee the same number of samples per second. Secondly, it can match mis-aligned trajectories by finding the optimal warping path which is important due to different sensitivities from different phone's sensor, whereas other distance-based measures (e.g. Euclidean, Manhattan) simply align the $i^{th}$ point on Alice's time series to the same $i^{th}$ point on Bob's time series (see Figure~\ref{euclideanalignment}). Thirdly, DTW is a proven technique with successfully time-tested applications in the speech recognition research community~\cite{senin2008dynamic,hu2003polyphonic}.

Lastly, our justification for using DDTW instead of the standard Dynamic Time Warping (DTW) is that DTW may suffer from incorrect alignments where a single position on Alice's trajectory is mapped onto a large set of positions on Bob's trajectory (see Figures~\ref{dtwalignment} and~\ref{dtwwarping}). This phenomenon commonly happens when standard DTW tries to explain the variability of the Y-axis by over-warping the X-axis (see Figure~\ref{euclideanvsdtw}).

Without loss of generality, given Alice's magnetic sequence $A=(m_1, \dots, m_N)$ and Bob's magnetic sequence $B=(m'_1, \dots, m'_M)$, DDTW first tries to build an N-by-M matrix, where the $[i^{th}, j^{th}]$ element is the distance between the two points $m_i$ and $m'_j$. While standard DTW uses the Euclidean distance, DDTW uses the square of the difference of the derivatives of $m_i$ and $m'_j$ as follows. This distance was empirically proven to be more robust to outliers than other estimate using only two data points~\cite{keogh2001derivative}.
\begin{equation}
D(A) = \frac{m_i - m_{i-1} + ((m_{i+1} - m_{i-1})/2)}{2}, (1 \leq i \leq M)
\end{equation}

\textbf{Step 4: Validating the matching pairs of trajectory}

Given one of Alice's trajectories, DDTW will always find a best matched trajectory from Bob's (i.e. the one with the smallest distance), although they may not be similar at all. This is a typical challenge for all distance-based and similarity-based approaches. For a highly sensitive problem such as epidemic tracking, an administrator normally looks at the final matching trajectories presented by the algorithm from the last step, and manually decides whether they are indeed co-located or not. Nevertheless, we present three heuristics to automate this decision-making process.
\begin{enumerate}
	\item The temporal difference of the two trajectories must be less than 5 seconds. For a typical 8-carriage train in London, it is unlikely that Alice and Bob are in the same carriage if their trajectories were distanced by more than a few seconds apart.
	\item The compression rate must not exceed 1.5. This number measures how stretched or compressed one trajectory is, in order to match the other trajectory. Realistically, we expect the journey of two co-located passengers to be roughly equal in terms of length. Given the length of Alice's magnetic trajectory is $l_A$ (samples) and Bob's is $l_B$ (samples), the compression rate is calculated as $\frac{max(l_A, l_B)}{min(l_A, l_B)}$.
	\item The difference score between the two trajectories must not exceed an empirical constant of 5. This score is calculated by adding up the difference between every aligned samples on the time series, divided by the total length of the warped path.	
\end{enumerate}

A pair of trajectories must satisfy all three above criteria to be declared as valid matching, and thus, signalling a co-location detection between the two respective passengers. We will evaluate their performances in the experimental section.

\subsection{Challenges to our approach}
Firstly, with any technique that aims to differentiate the users' position, the spatial uniqueness of the sensor reading is essential. Although our approach takes into account the time series of the sensors' reading, if the user takes a very short trip, it is much harder to match his trajectory to other passengers'. We will assess this challenge in the experimental section.

Secondly, time-wise, all users' phones must be synchronised to correctly co-locate their owners. Since the app uses the local time of the phone to stamp each sensor output, some mismatch between different phones' clock may occur. A simple solution is to inquire an internet time service or the cellular provider for ground-truth, whenever a connection can be made. This ground-truth will help revealing the offset to the phone's local time.

Thirdly, the heterogeneous devices remain a difficult task for any smart phone based approach. Different models may employ non-identical chip sensors, which have different sensitivities. However, our algorithm does not consider the absolute strength value, but looks at the overall shape of the trajectories to match them.

\section{Empirical experiments}
This section conducts the experiments to assess the feasibility and the accuracy of our approach. In doing so, it aims to address the following research questions.

In terms of feasibility:
\begin{itemize}
	\item \textbf{How much spatial variation does the on-board magnetism possess?} High variation of magnetism amongst places is highly desirable to generate a distinguishable trajectory for people in different carriages.
	\item \textbf{How identical is the magnetic field strength in the same train carriage or bus?} We hypothesise that nearby passengers at carriage-level should observe a similar magnetic reading at any moment.
\end{itemize}

In terms of accuracy:
\begin{itemize}
	\item \textbf{What is the precision and recall rate of our co-location detection algorithm?} We will verify the successfulness of our detection algorithm on real-world data.
\end{itemize}

\begin{figure}
	\centering
	
	\subfloat[Overground train test routes. They cover over 70 km, passing through 2 of the busiest stations in London (London Bridge \& Liverpool Street).]{\includegraphics[width=2.5in]{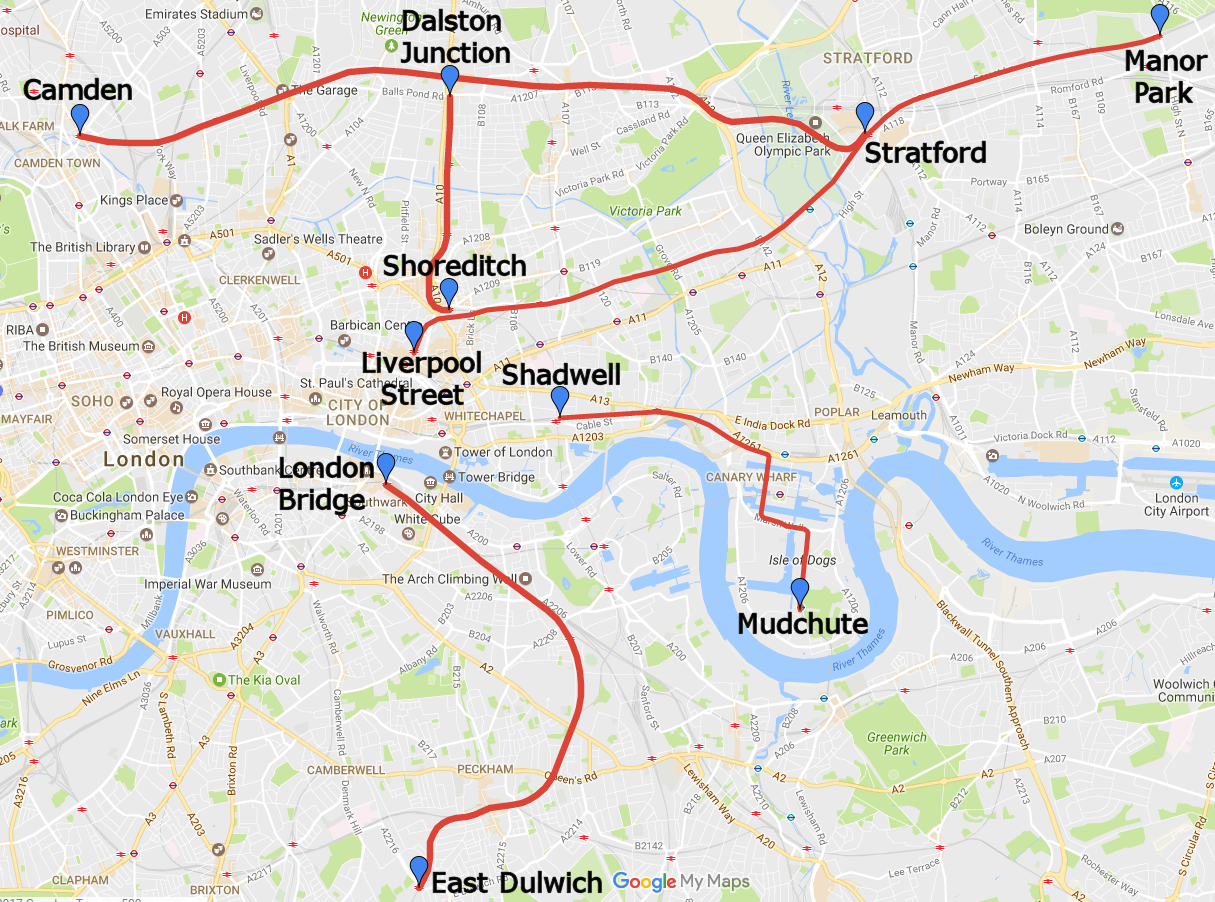}
		\label{overgroundtest}}
	\hfil
	\subfloat[Underground tube test routes. The routes shown here are exactly the same as the real-life ones. Since the tubes travel underground, some paths appear to go under-water and through buildings.]{\includegraphics[width=2.5in]{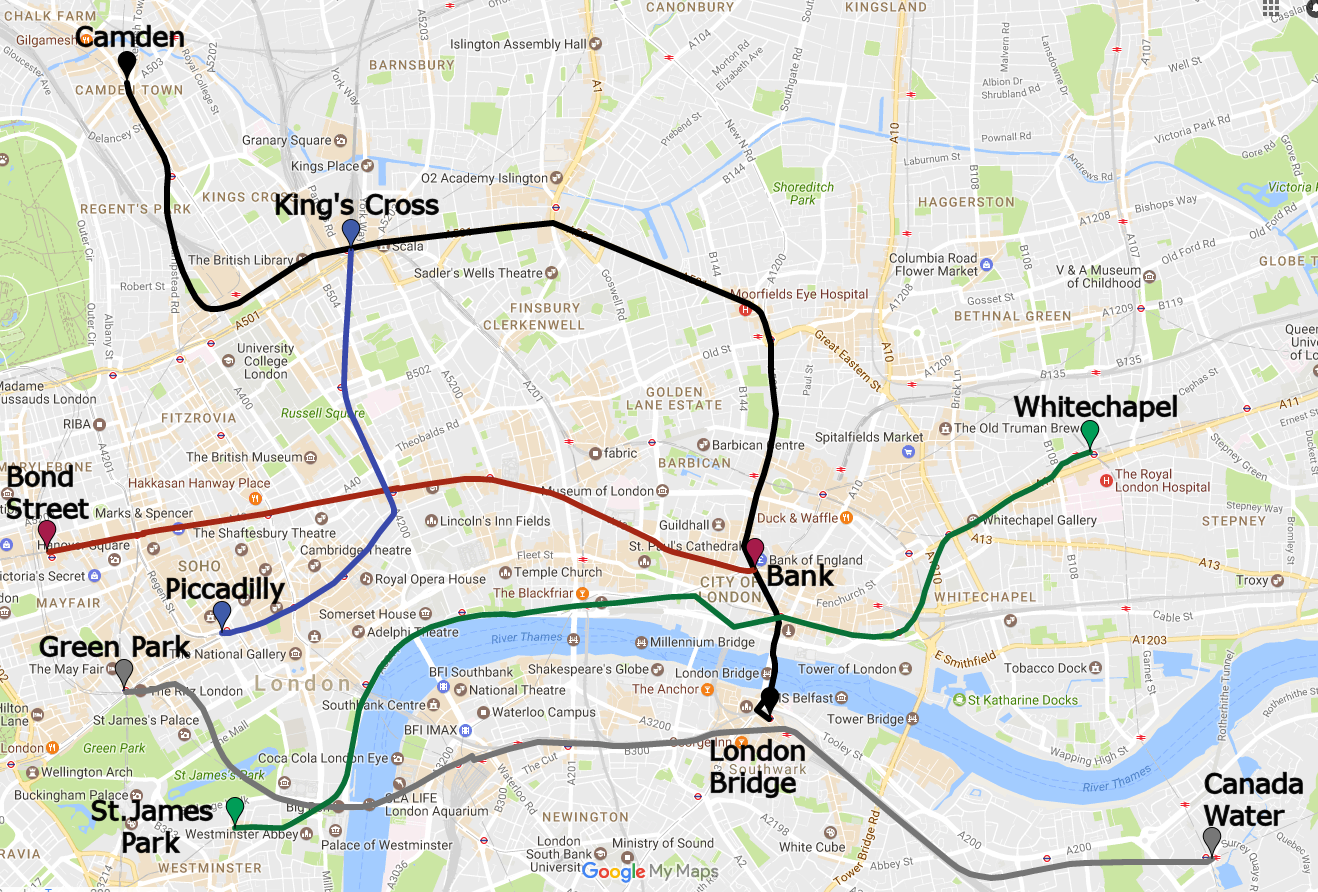}
		\label{undergroundtest}}
	\hfil
	
	\caption{The overground and underground test environments visualised on Google Maps.}
	\label{overundertest}
\end{figure}

\begin{figure*}
	\centering
	
	\subfloat[East Dulwich - Peckham Rye route.]{\includegraphics[width=2.0in]{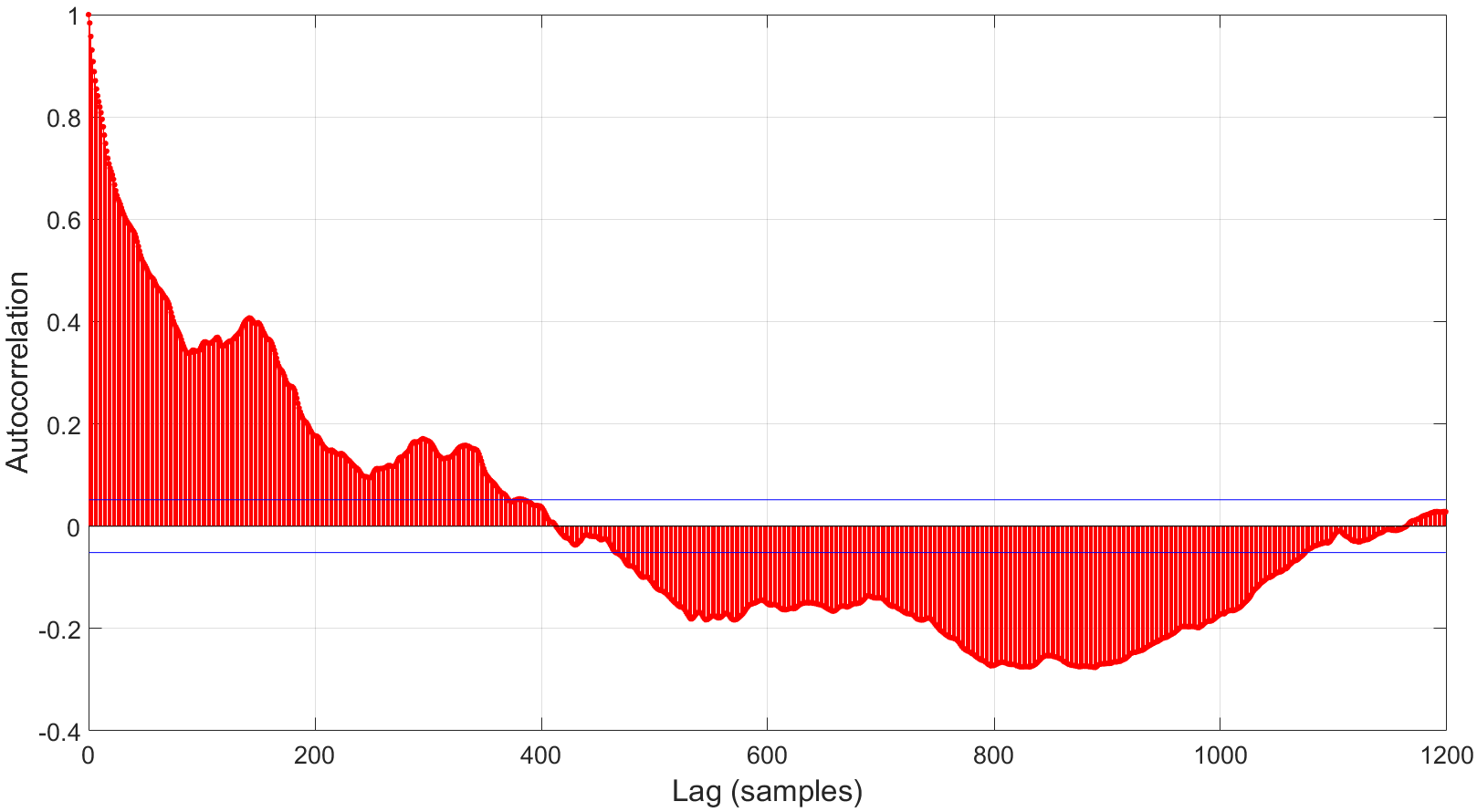}
		\label{overgroundautocorrelation1}}
	\hfil
	\subfloat[Peckham Rye - Queens Road route.]{\includegraphics[width=2.0in]{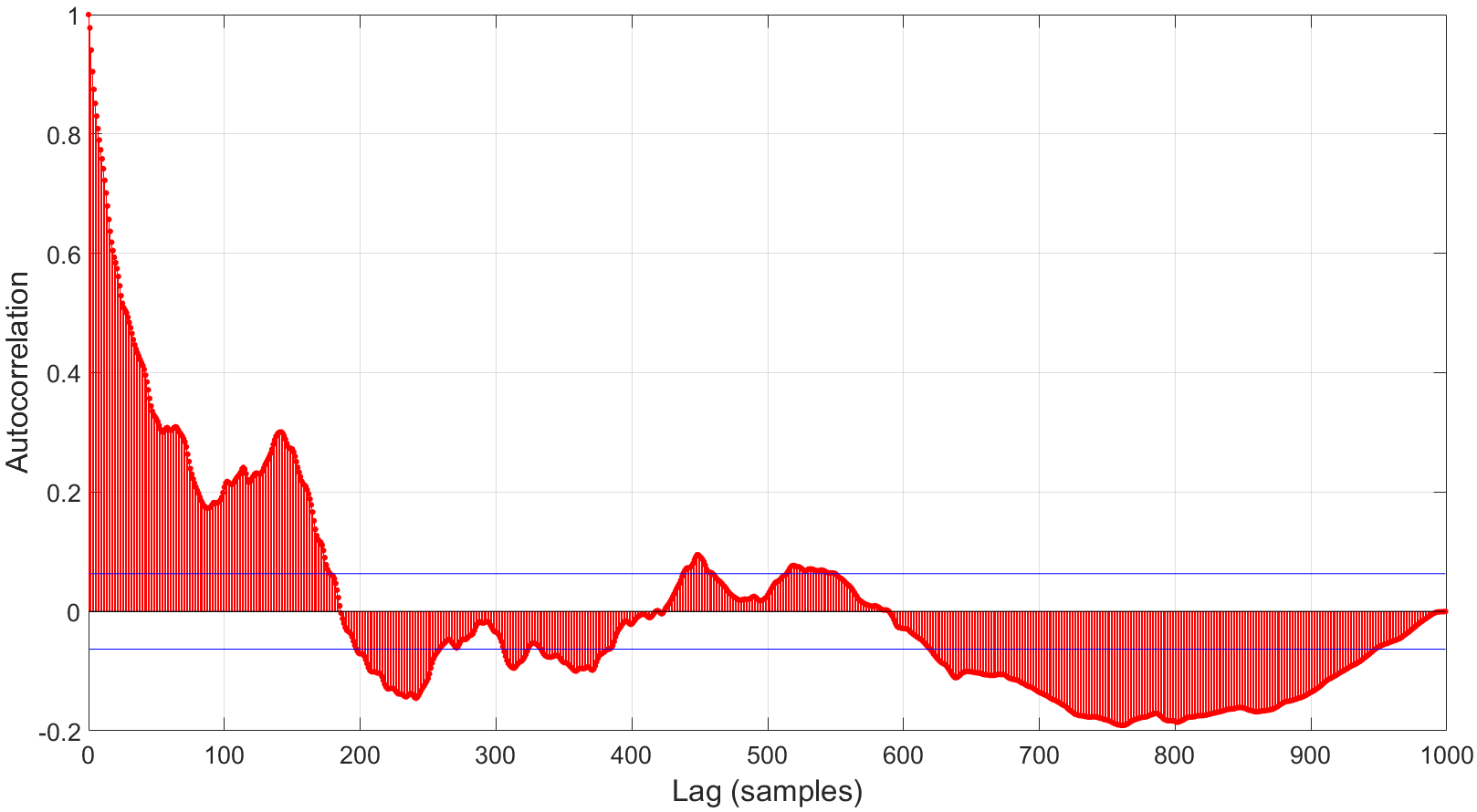}
		\label{overgroundautocorrelation2}}
	\hfil
	\subfloat[Queens Road - South Bermondsey route.]{\includegraphics[width=2.0in]{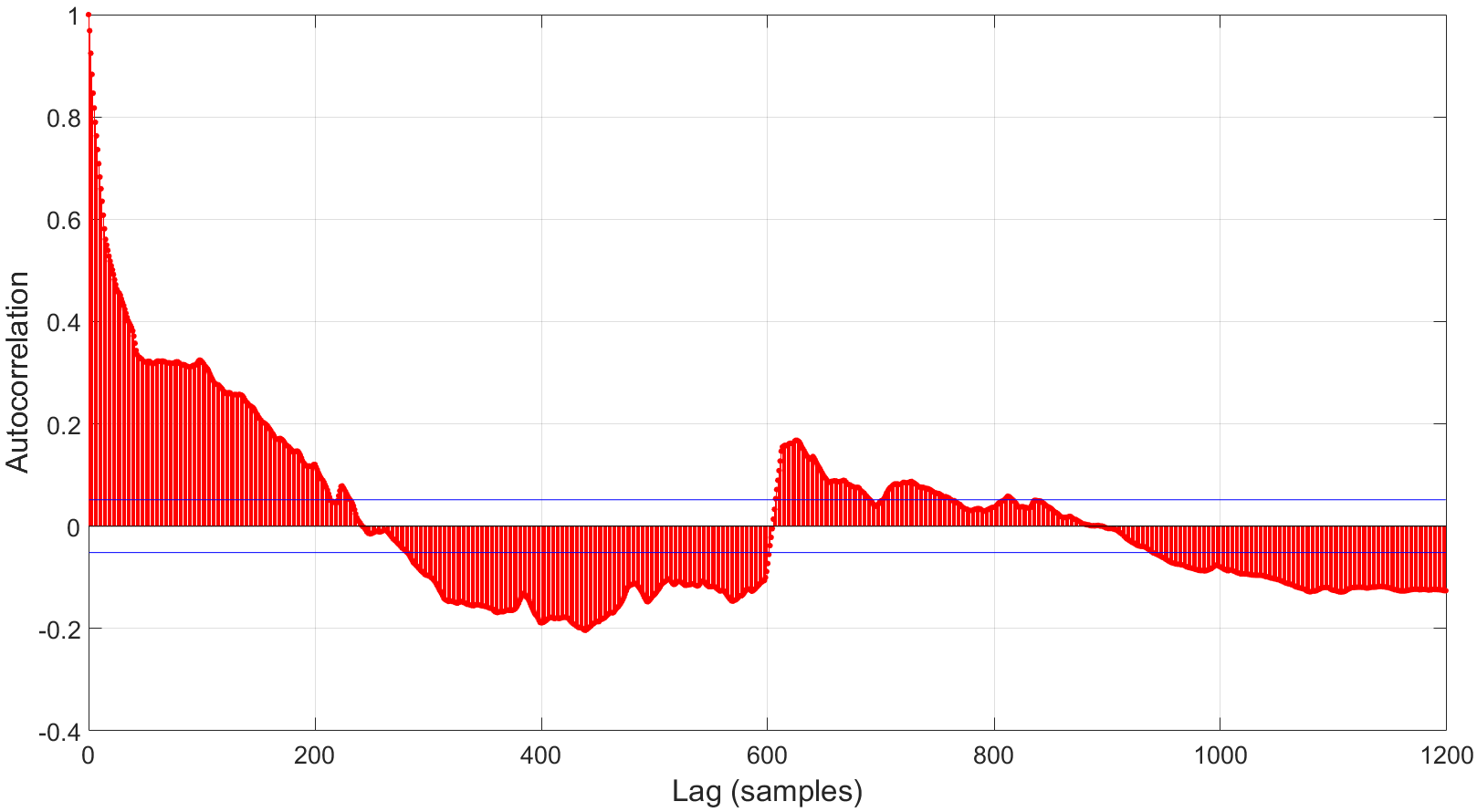}
		\label{overgroundautocorrelation3}}
	\hfil
	
	\subfloat[Covent Garden - Holborn route.]{\includegraphics[width=2.0in]{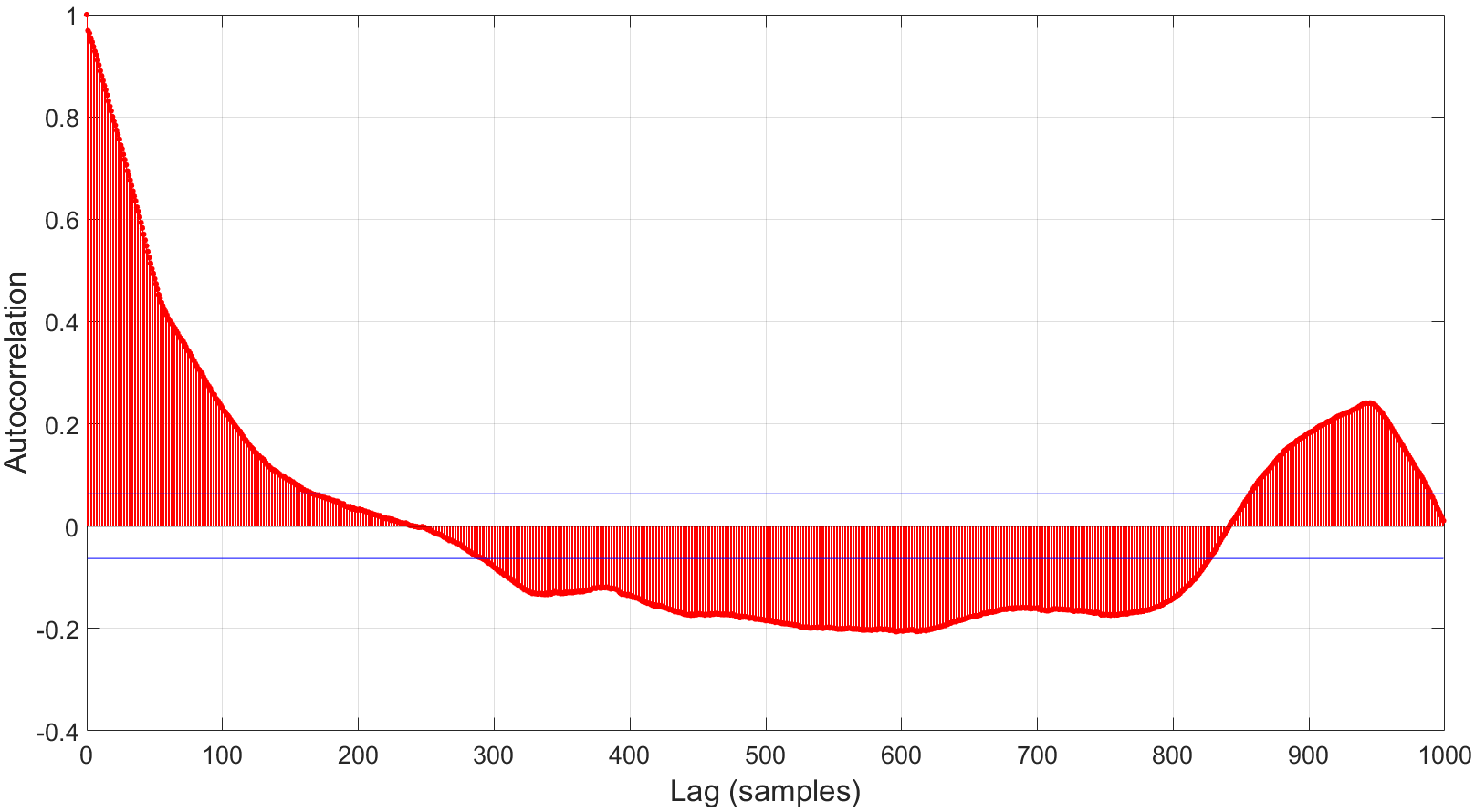}
		\label{undergroundautocorrelation1}}
	\hfil
	\subfloat[Old Street - Angel route.]{\includegraphics[width=2.0in]{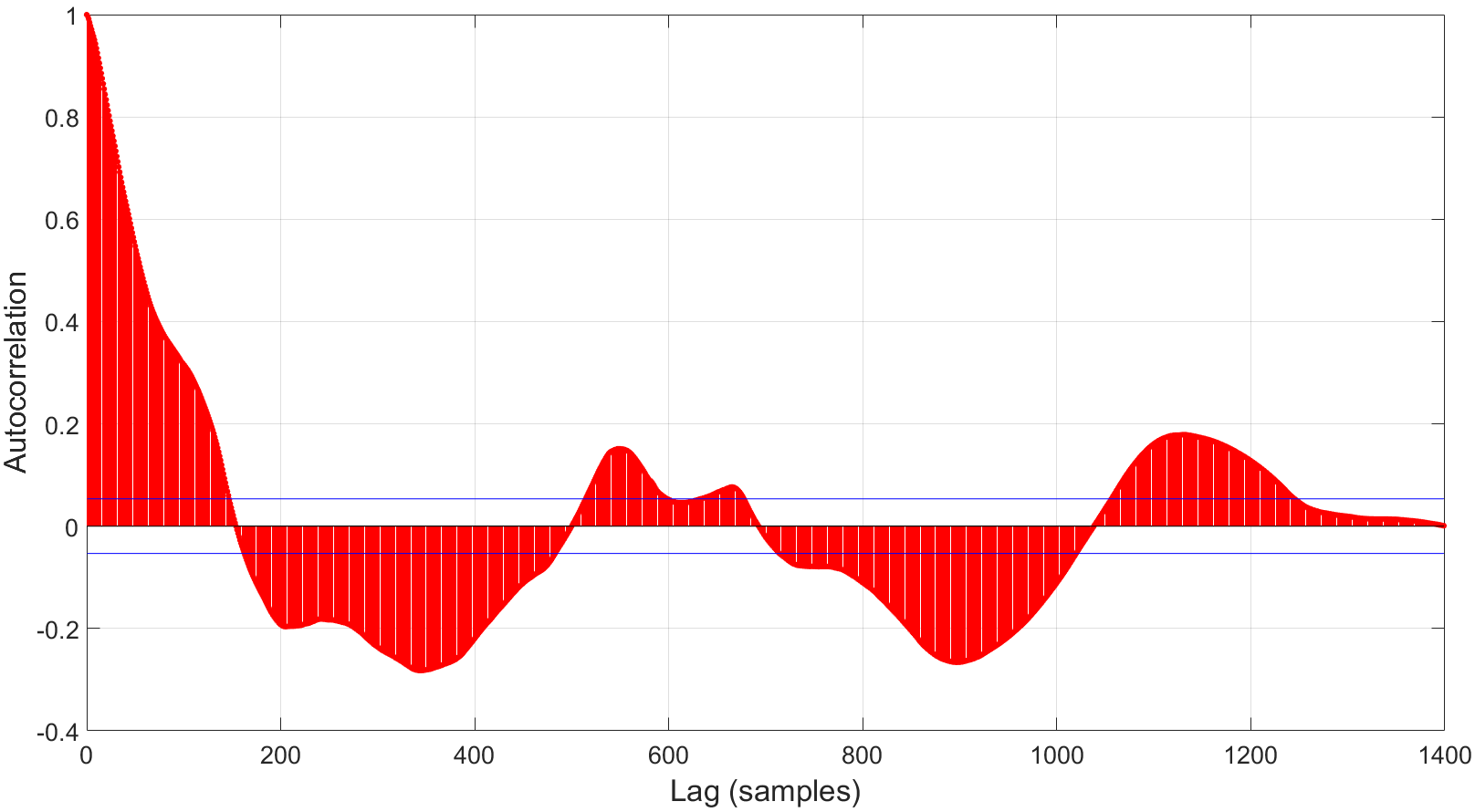}
		\label{undergroundautocorrelation2}}
	\hfil
	\subfloat[Waterloo - Southwark route.]{\includegraphics[width=2.0in]{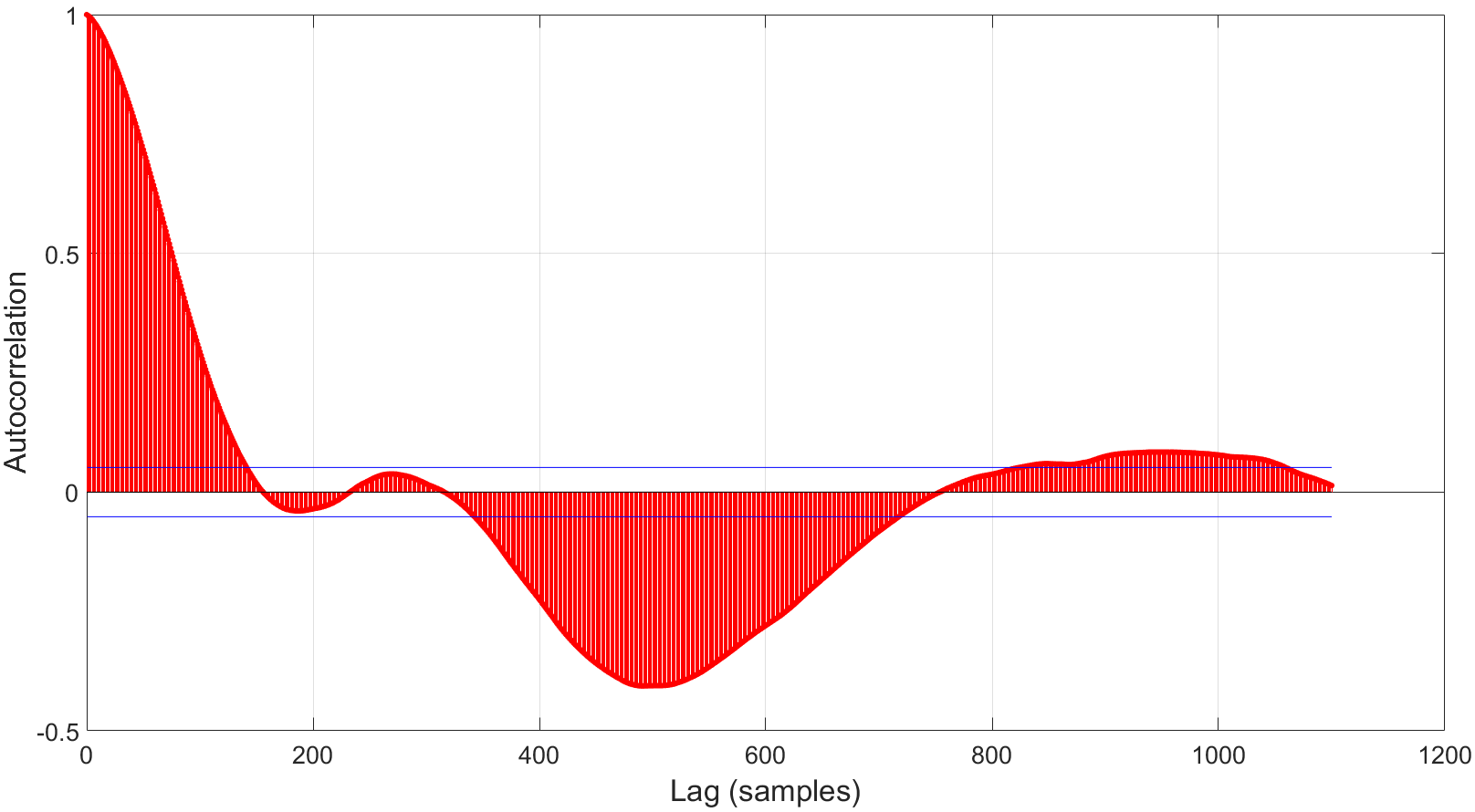}
		\label{undergroundautocorrelation3}}
	\hfil
	
	\caption{The autocorrelation plot of the six test trajectories. We omit the remaining trajectories which exhibits a similar trend for page limit. The majority of autocorrelations are non zero, which confirms the non-stationary property of these magnetic time series.}
	\label{autocorrelation}
\end{figure*}

For the ease of assessment, the experiments were separated into three categories - overground train, underground tube and the bus, which cover diverse landscapes of London. Different types of vehicles were also tested (i.e. London trains are operated by 22 different companies\footnote{http://www.londontravelwatch.org.uk/links/train\_operating\_companies - last accessed in Feb/2017}). Two Android phones were used in this research, namely the Google Nexus 5 (released in 2013, running Android Lollipop), and Lenovo Phab 2 Pro (released in 2016, running Android Marshmallow). Through-out the following experiments, these devices were held naturally in the surveyor' hands, or left in the pocket. Their local clocks are also synchronised.

\subsection{Overground train and underground tube test environments}
As the overground trains and underground tubes share similar aspects (i.e. both have multiple carriages, are electric-based), we combined both test environments for more concise analysis.

Our overground test environment is made of 5 separate routes, which traverses 31 different stations, and covers over 70 kilometres of travelling distance in the South-East and East-Central of London (East Dulwich - London Bridge - Camden - Liverpool Street - Stratford - Manor Park) (see Figure~\ref{overgroundtest}). Our underground tube test scenario examines 5 main lines of the London underground network, namely the Northern, Central, Jubilee, Piccadilly and District line, covering over 57 kilometres (see Figure~\ref{undergroundtest}). For both test environments, each route was visited twice with the surveyors in different seats and carriages. We used 4 different train companies to add more diversities to the dataset.

The first experiment assesses the spatial variation of the on-board magnetism. A surveyor sat in the same place and travelled through all of the above test routes. We then examine the resulting magnetic trajectory between every 2 consecutive stations on his journey. Our hypothesis is that all trajectories are non-stochastic or non-stationary (i.e. we want the magnetic field strengths within a trajectory to change significantly). Visually speaking, an autocorrelation plot of each trajectory time series has significant non-zero lags, which confirms the trajectory is non-stochastic. Additionally, the line segment's length gradually decreases below zero, which indicates a non-stationary time series (see Figure~\ref{autocorrelation}).

\begin{figure}[!t]
	\centering
	
	\subfloat[Waiting at Startford (overground) on Carriage \#1.]{\includegraphics[width=1.7in]{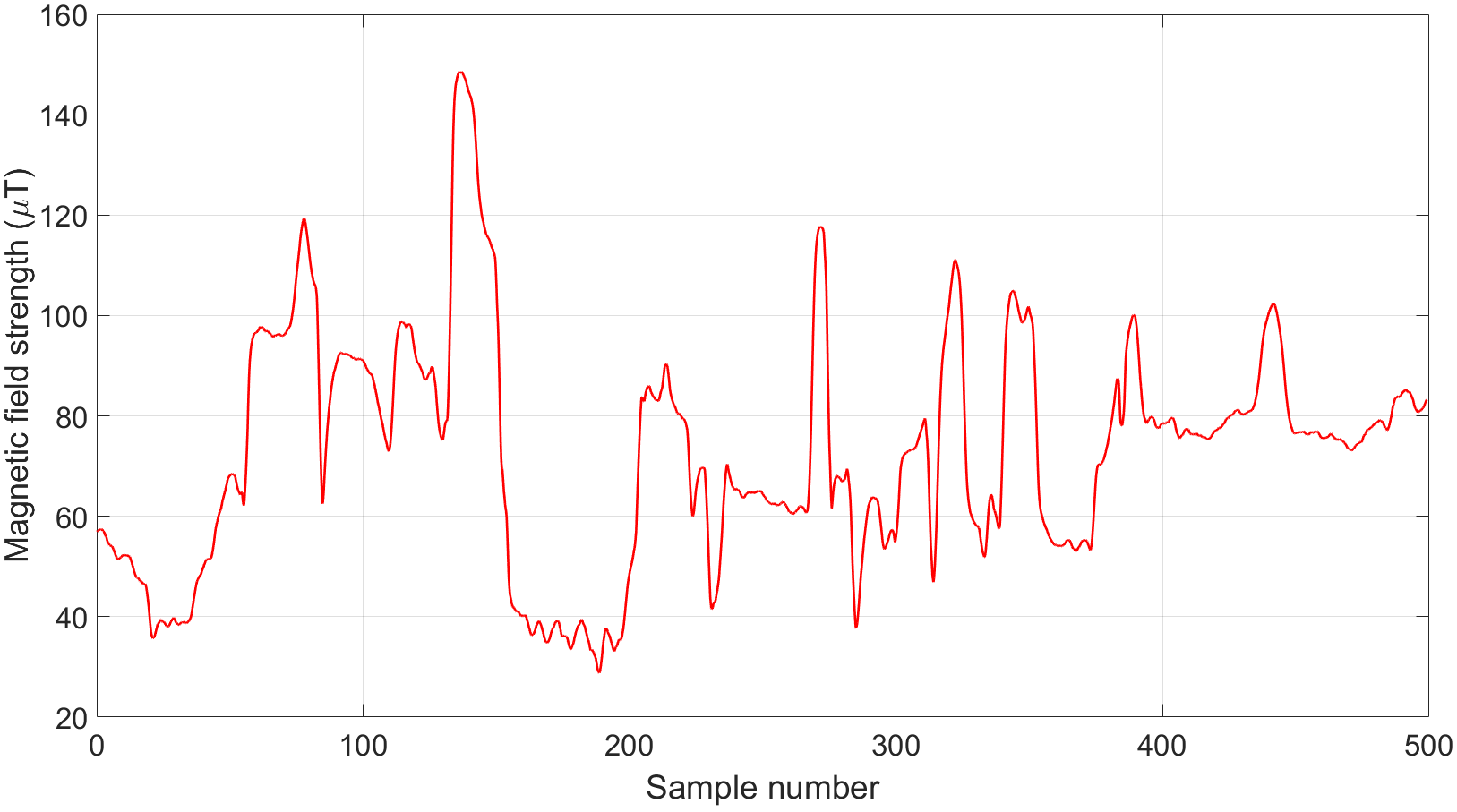}
		\label{waitingatstation1}}
	\subfloat[Waiting at Startford (overground) on Carriage \#4.]{\includegraphics[width=1.7in]{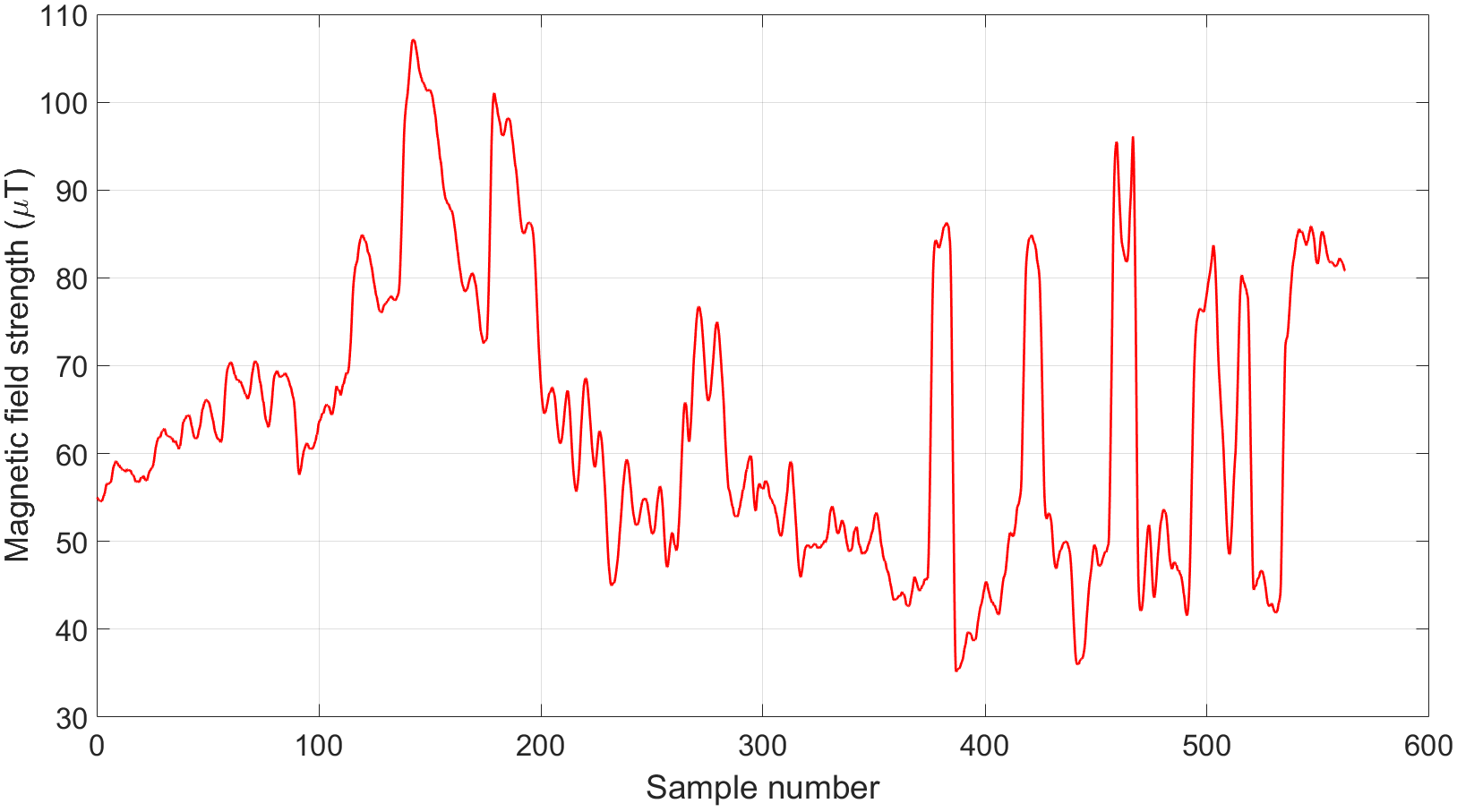}
		\label{waitingatstation2}}	
	
	\subfloat[Waiting at Holborn (underground) on Carriage \#1.]{\includegraphics[width=1.7in]{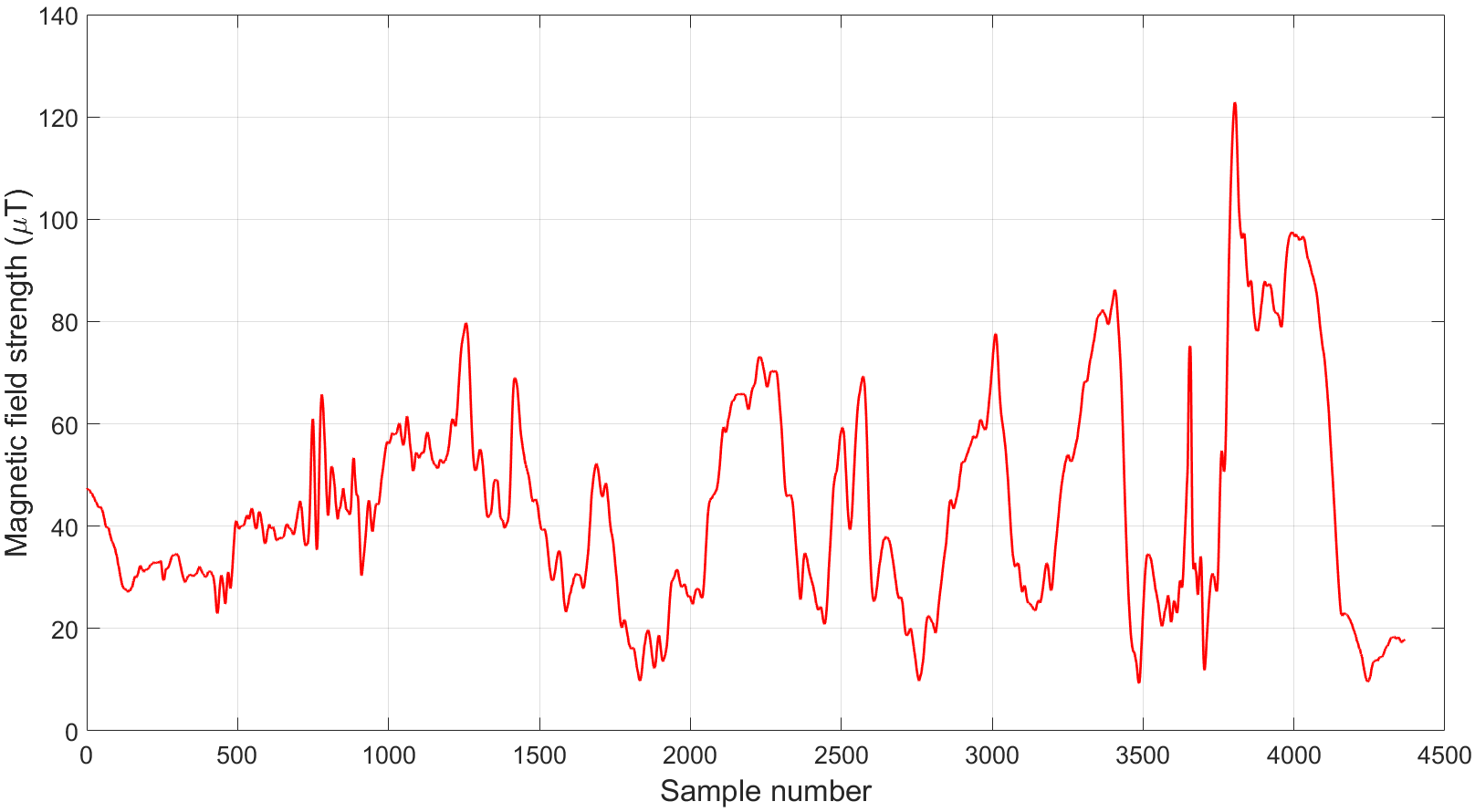}
		\label{waitingunderground2}}
	\subfloat[Waiting at Holborn (underground) on Carriage \#3.]{\includegraphics[width=1.7in]{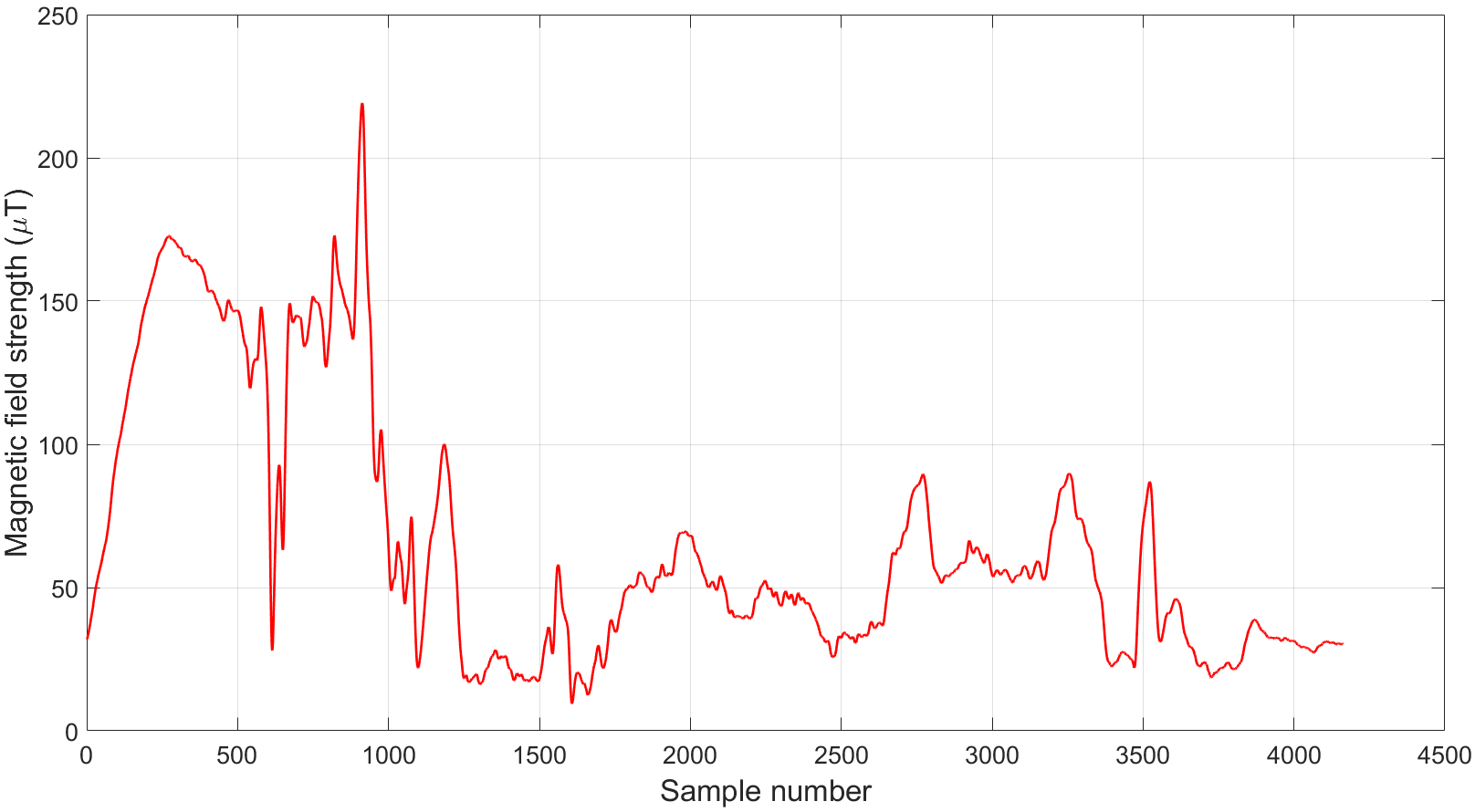}
		\label{waitingunderground3}}		
	
	\caption{The on-board magnetism readings from 2 different carriages on a static train. No movement from either surveyor or nearby passengers existed. This experiment proved the strong non-uniform impact of the electric current from the railway structure on different train carriages.}
	\label{waitingatstation}
\end{figure}

What surprised us the most when carrying out this experiment was that often when the train waited at the station, the magnetometer reported high measures without any movement from both the user and the train. This phenomenon happened even at relatively quiet stations without much movements from other passengers on the platform. This ascertains our aforementioned assumption that electric-based trains greatly distort the on-board magnetic field. However, we discovered that not all carriages experienced the same effect (see Figure~\ref{waitingatstation}). This is a significant attribute for our purpose, since it combines with the natural magnetism distortion from nearby building structure to make the magnetic observations more unique.
\begin{figure*}[h]
	\centering
	
	\subfloat[East Dulwich - London Bridge route]{\includegraphics[width=2.0in]{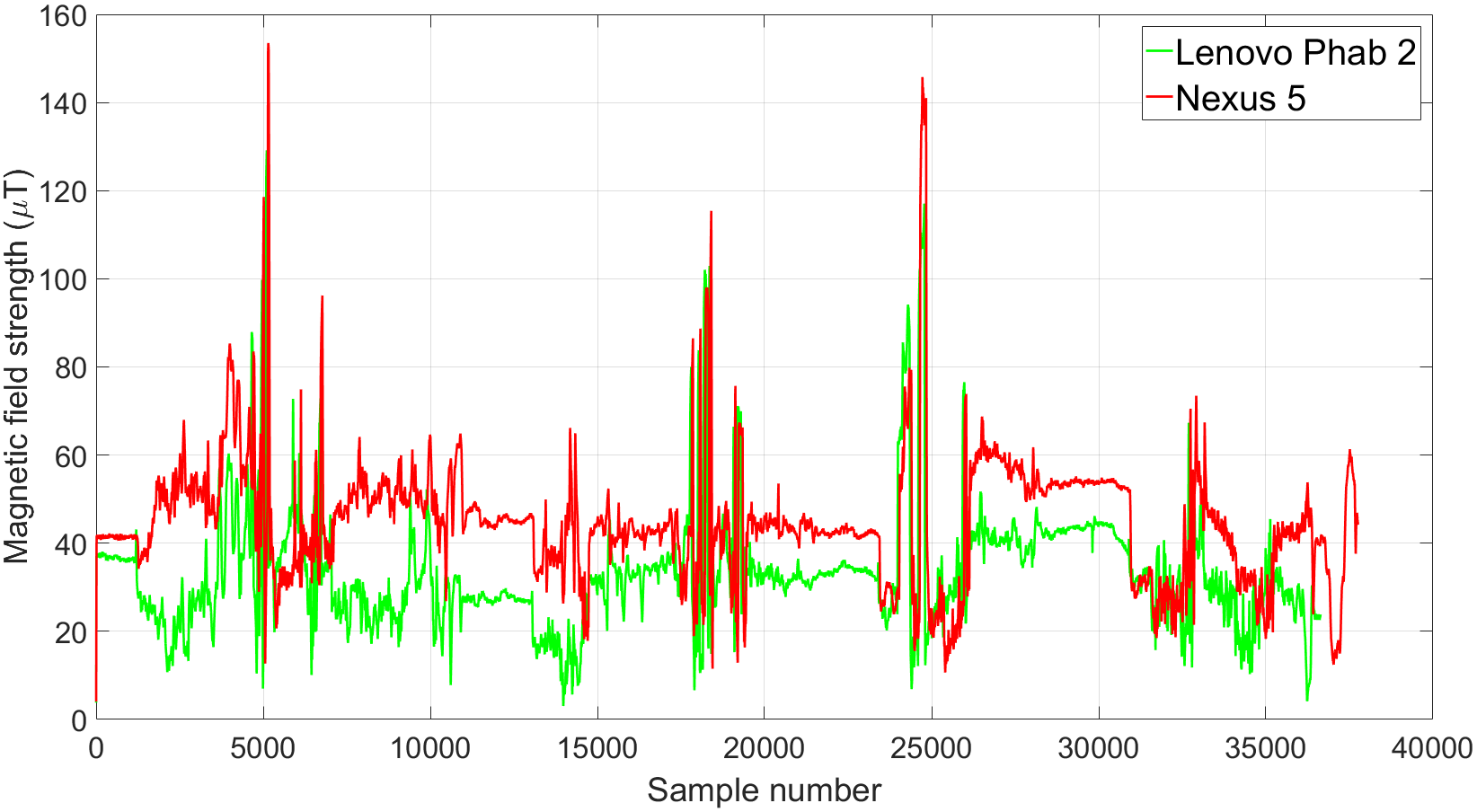}
		\label{overgroundsamecar1}}
	\hfil
	\subfloat[Camden - Stratford route]{\includegraphics[width=2.0in]{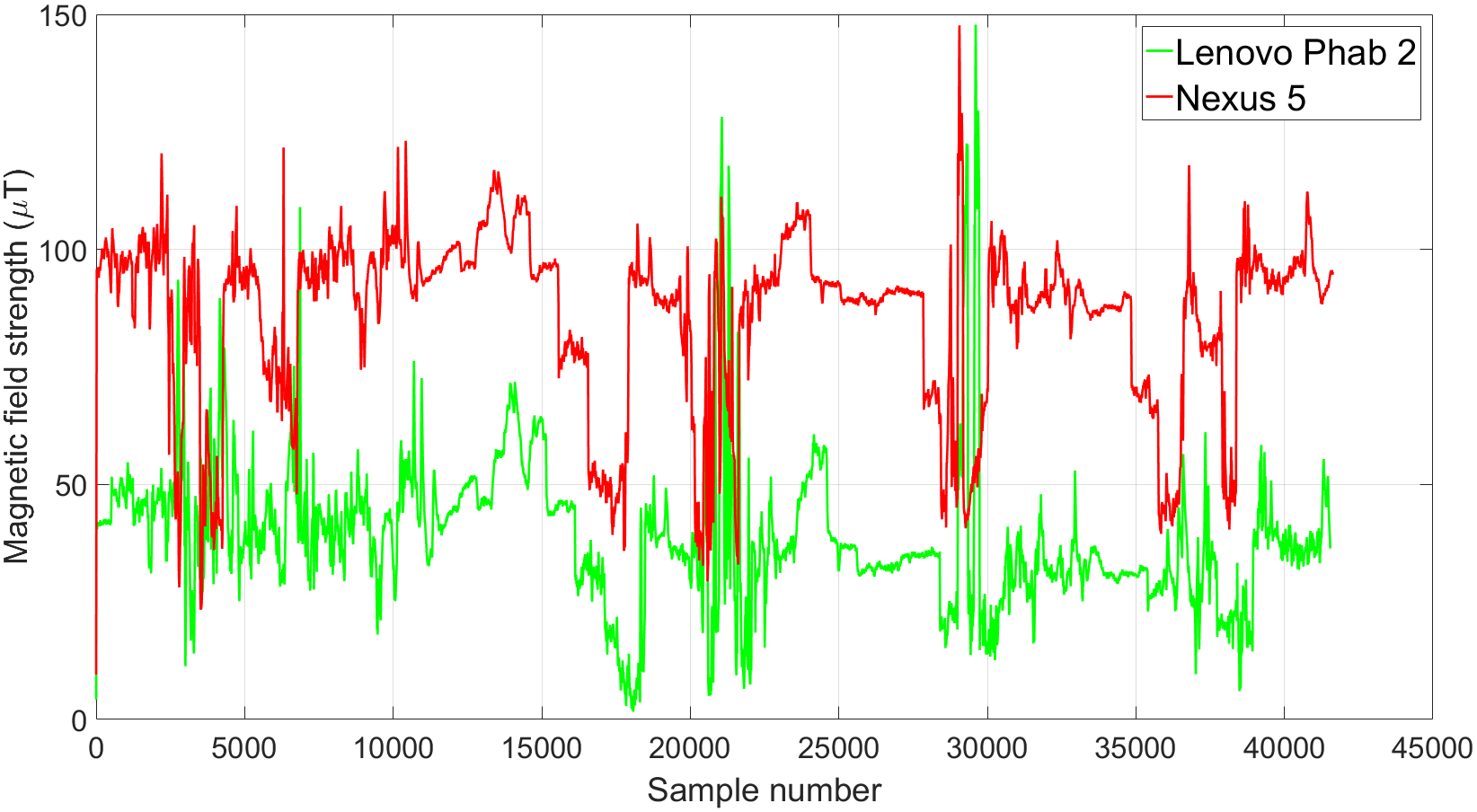}
		\label{overgroundsamecar2}}
	\hfil
	\subfloat[Shoreditch - Dalston Junction route]{\includegraphics[width=2.0in]{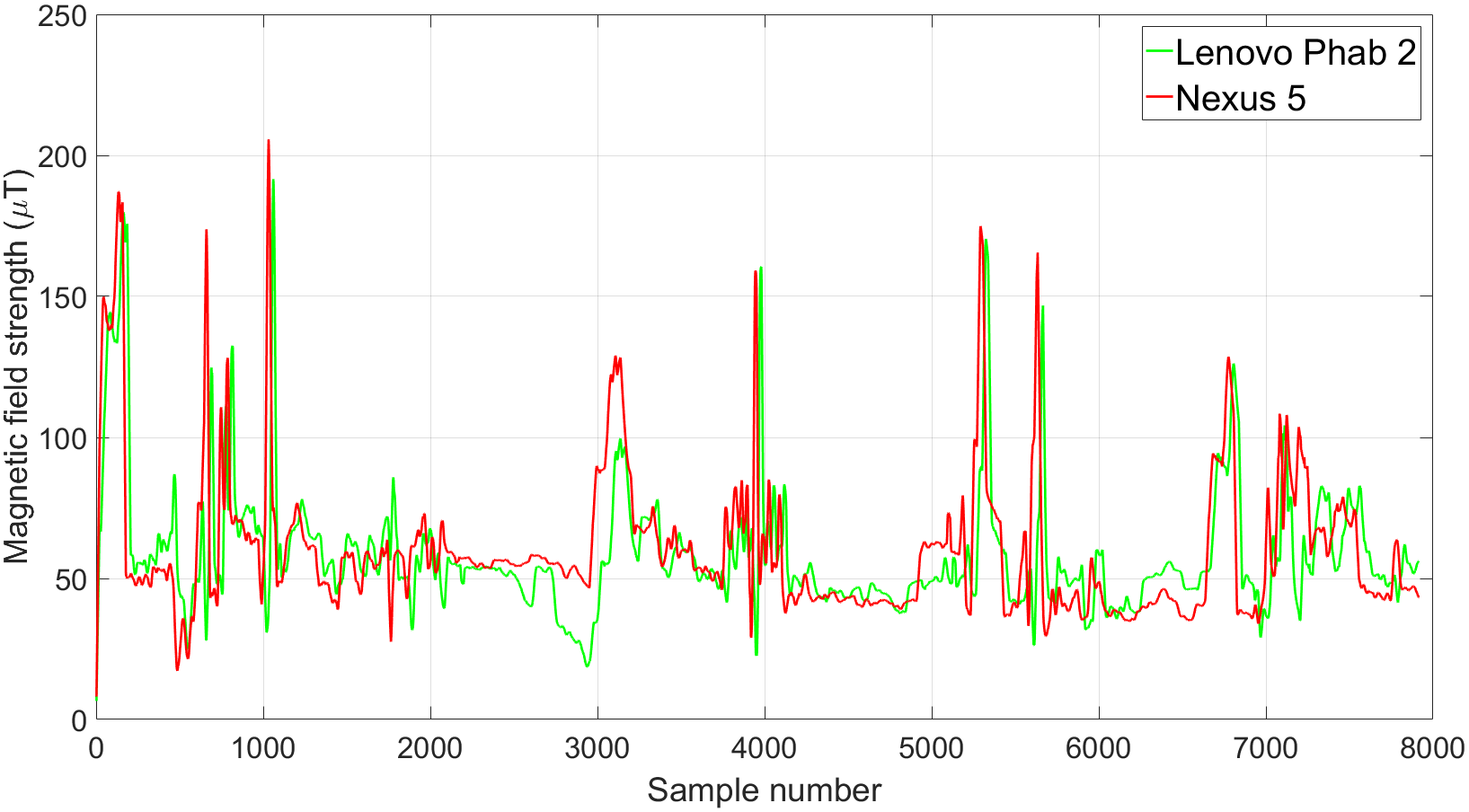}
		\label{overgroundsamecar3}}
	\hfil
	\subfloat[Shadwell - Mudchute route]{\includegraphics[width=2.0in]{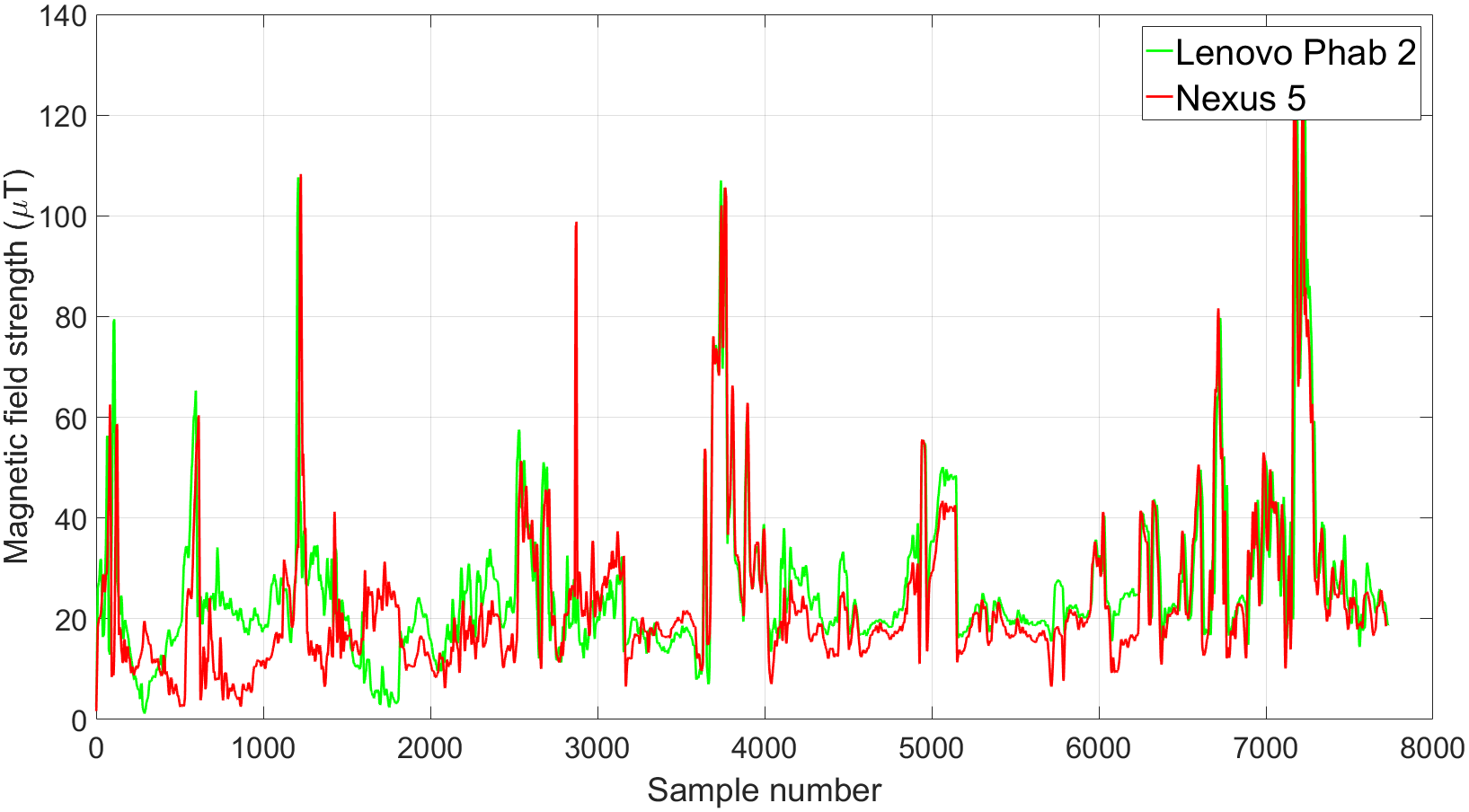}
		\label{overgroundsamecar4}}
	\hfil
	\subfloat[Liverpool Street - Manor Park route]{\includegraphics[width=2.0in]{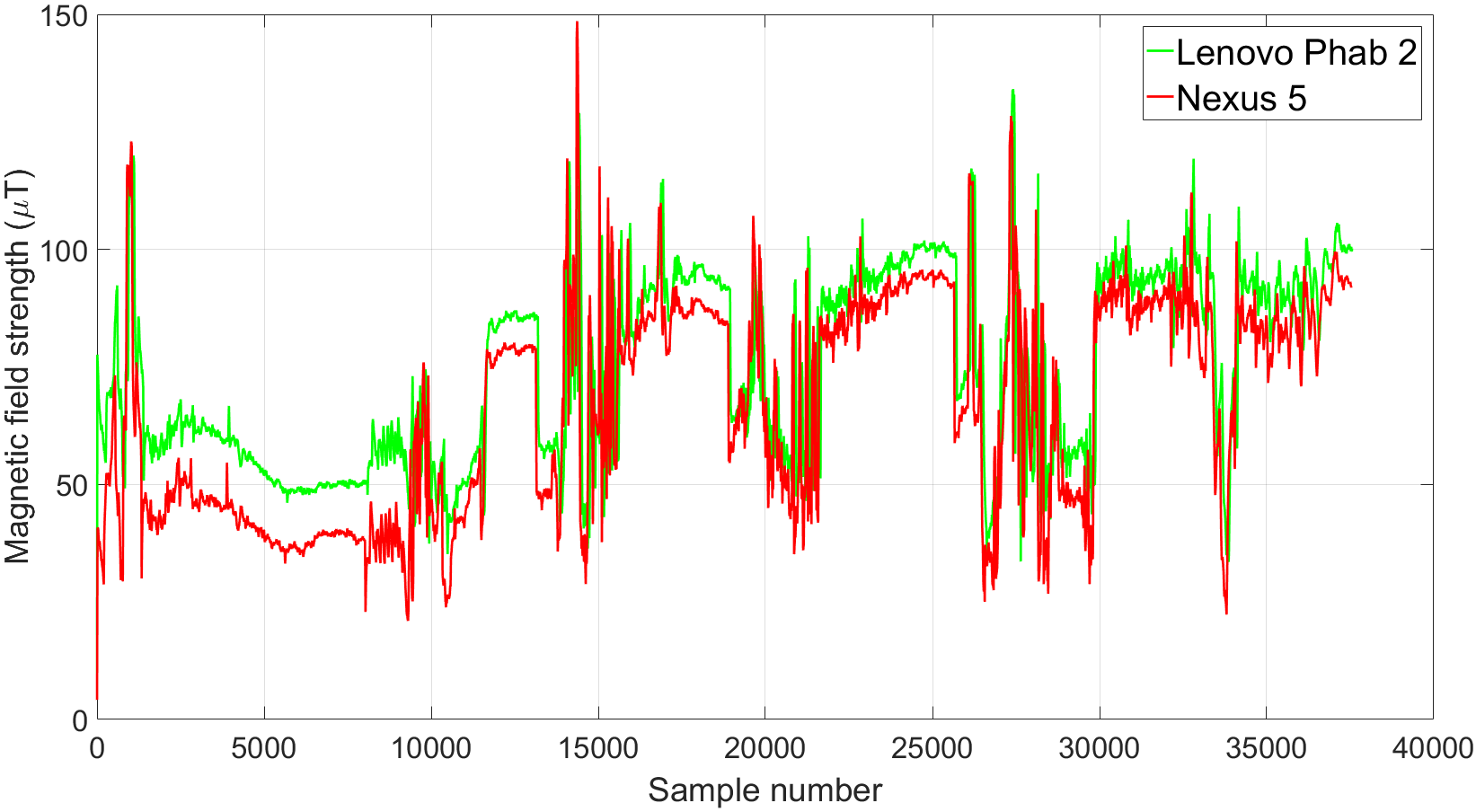}
		\label{overgroundsamecar5}}
	\hfil	
	
	\caption{The magnetic field observed by two mobile devices on the same carriage. All test trips exhibit a remarkably similar shape. The gap in the magnitude was caused by slightly different sensitivities from different phone models.}
	\label{overgroundsamecaroverground}
\end{figure*}
\begin{figure*}[h]
	\centering
	
	\subfloat[London Bridge - Camden (Northern line)]{\includegraphics[width=2.0in]{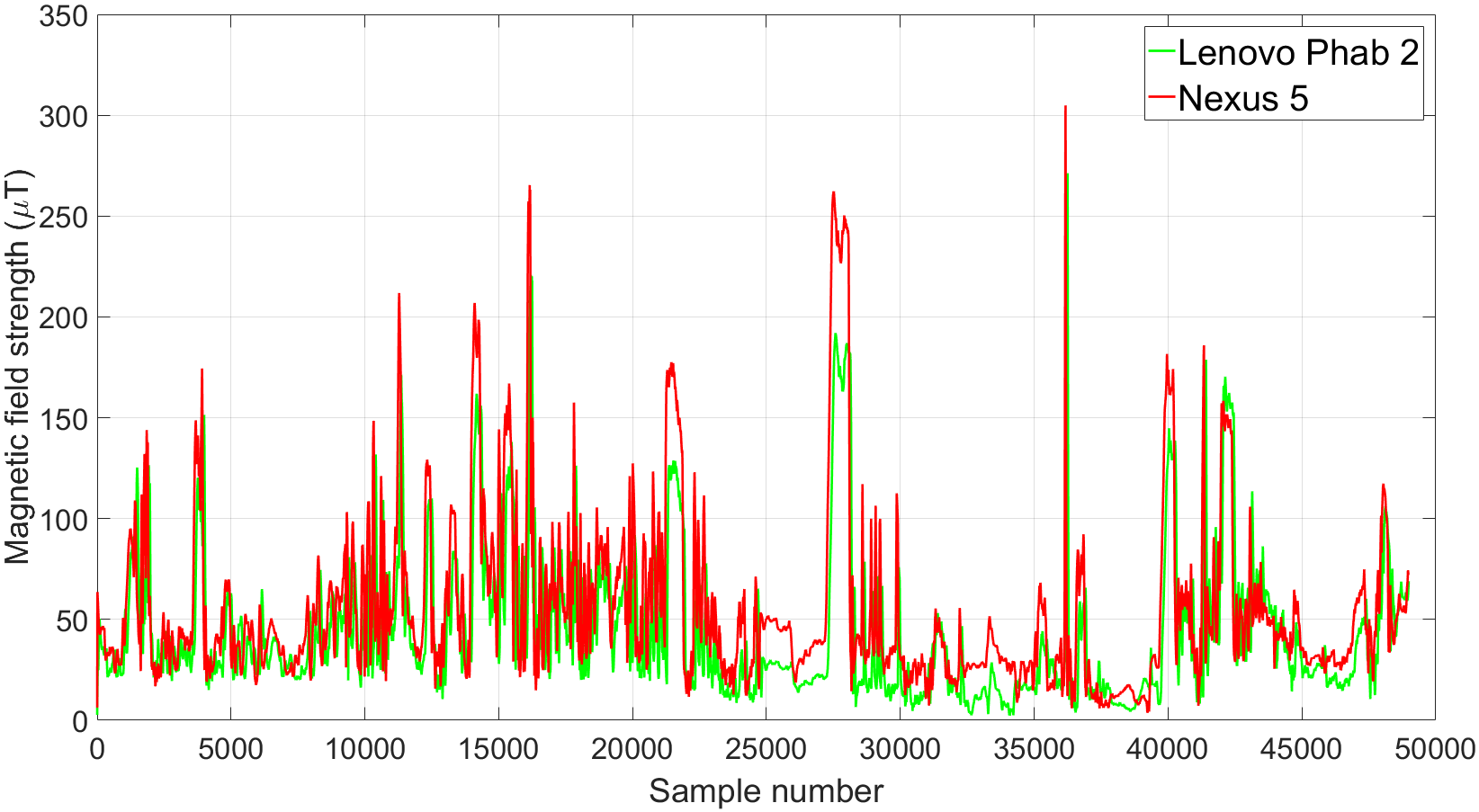}
		\label{undergroundsamecar1}}
	\hfil
	\subfloat[St. James' Park - Whitechapel (District line)]{\includegraphics[width=2.0in]{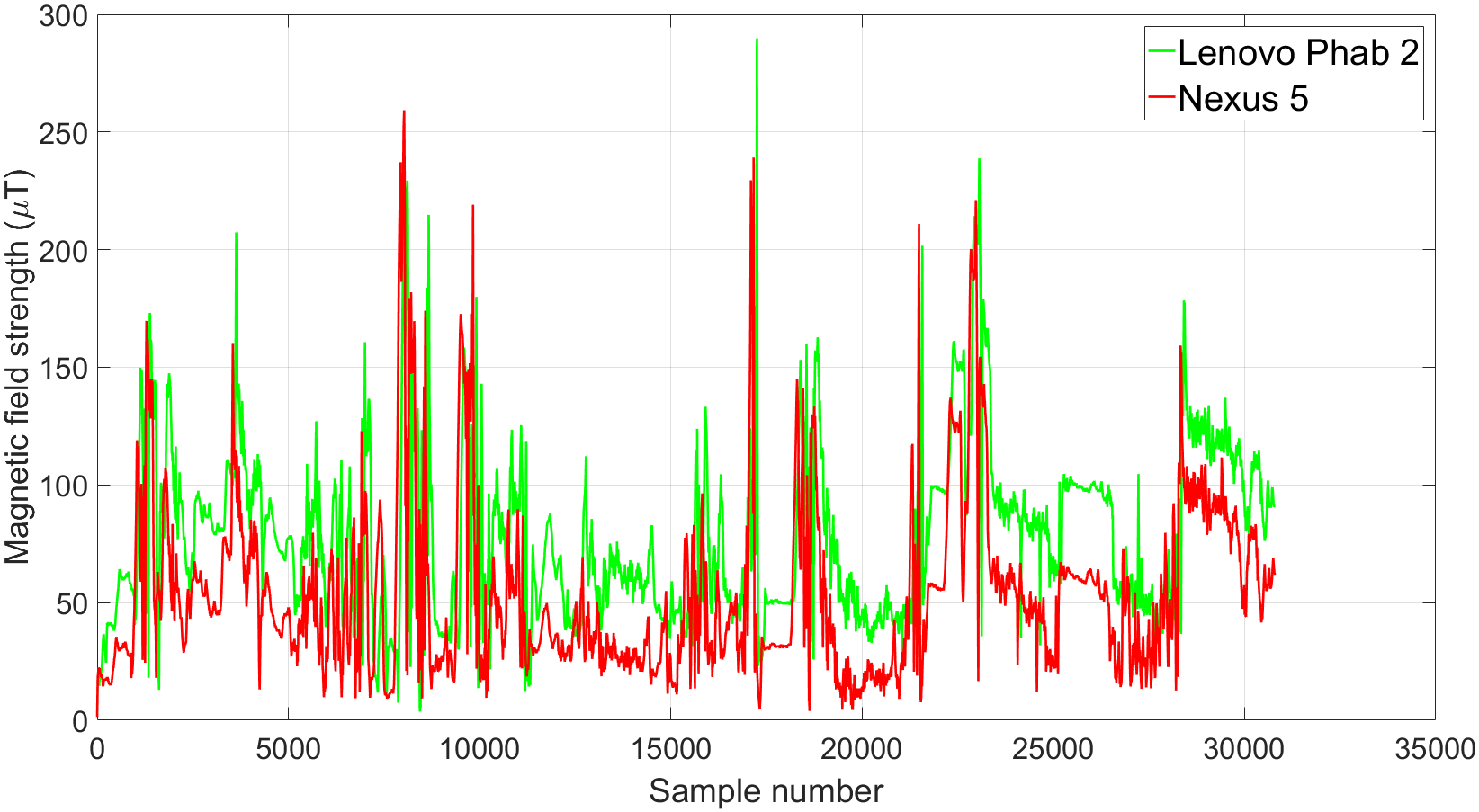}
		\label{undergroundsamecar2}}
	\hfil
	\subfloat[Green Park - Canada Water (Jubilee line)]{\includegraphics[width=2.0in]{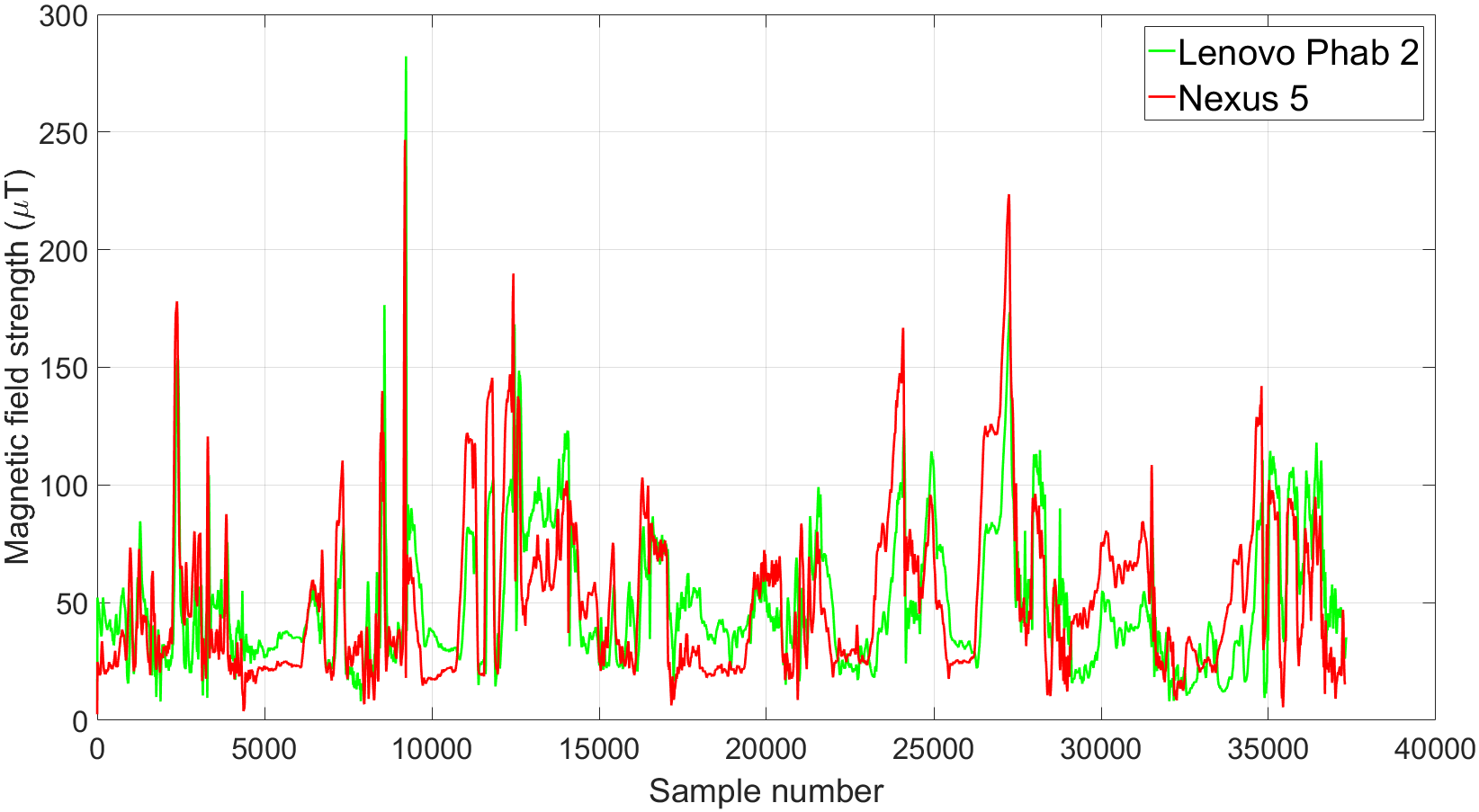}
		\label{undergroundsamecar3}}
	\hfil
	\subfloat[Piccadilly - King's Cross (Piccadilly line). The flat line was caused by non-moving tube because of congestion near Leicester Square station. It was interesting to observe that the magnetic reading stays relatively stable during this period]{\includegraphics[width=2.0in]{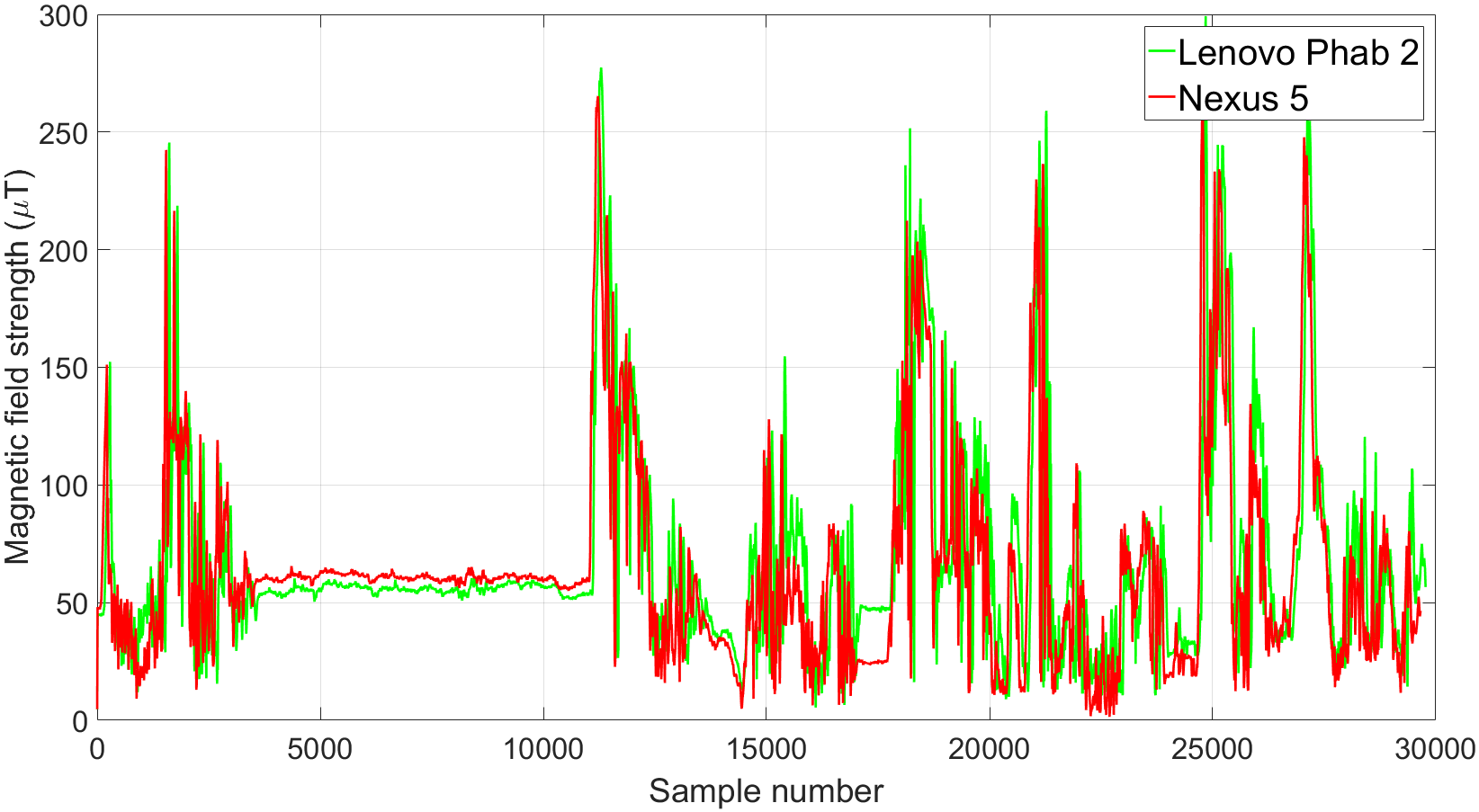}
		\label{undergroundsamecar4}}
	\hfil
	\subfloat[Bond Street- Bank (Central line)]{\includegraphics[width=2.0in]{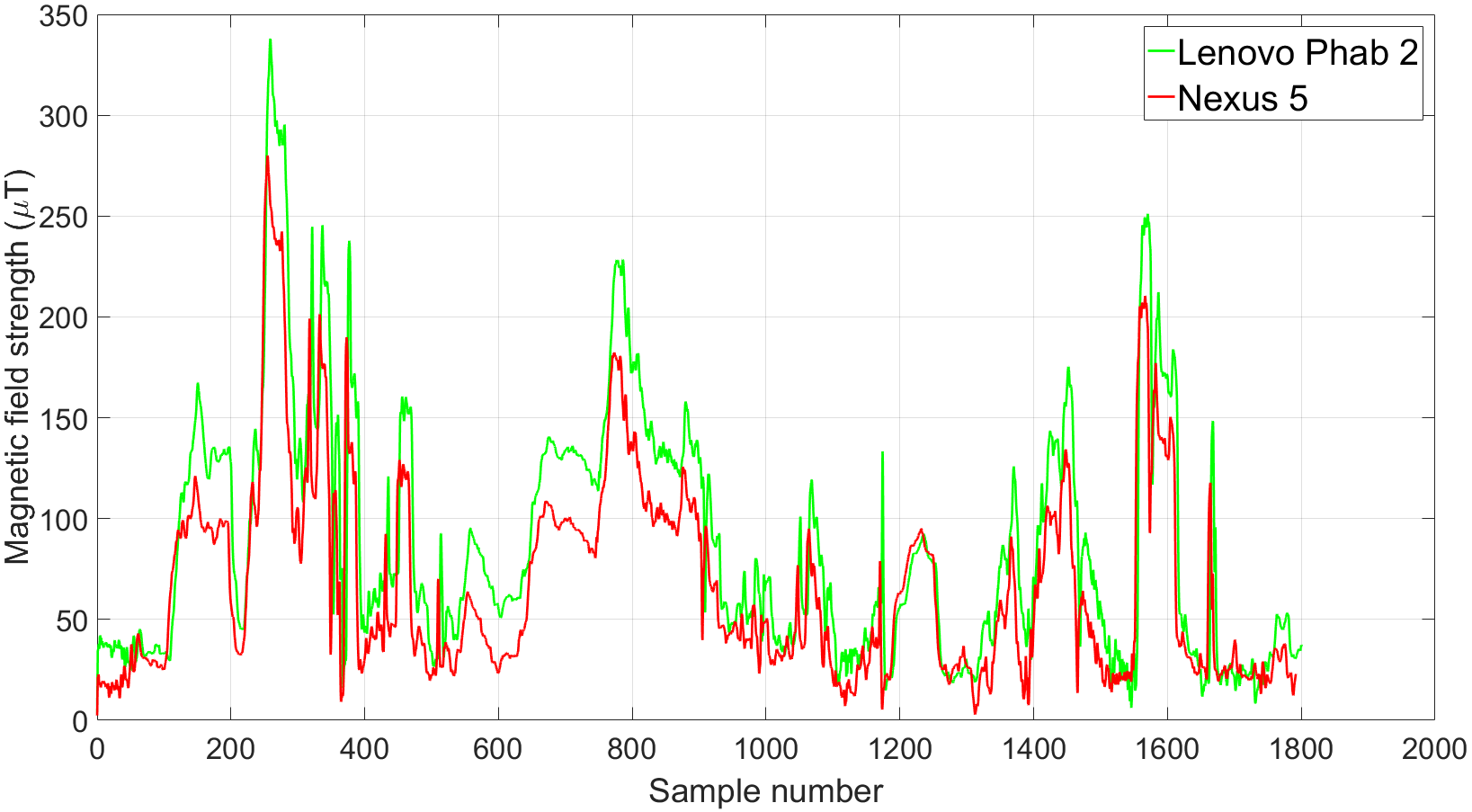}
		\label{undergroundsamecar5}}
	\hfil	
	
	\caption{Comparison of the underground trips observed by two mobile device on the same carriage. All test trips exhibit a remarkably similar shape. The gap in the magnitude was caused by slightly different sensitivities from different phone models.}
	\label{overgroundsamecarunderground}
\end{figure*}

The second experiment assesses the magnetic field strength observed by people on the same carriage. Our hypothesis is that their mobile devices should capture similar magnetism readings. For each trip, two surveyors sat on the same carriage, albeit in different seats. The maximum distance between them was up to 7 metres. Figures~\ref{overgroundsamecaroverground} and~\ref{overgroundsamecarunderground} display a remarkably similar shape of the two magnetic trajectories. The oscillation happened noticeably more often on the underground trips than the overground ones.

\begin{figure}[!t]
	\centering
	
	\subfloat[The DDTW scores heuristic comfortably rejected all non co-located pairs with fine margins.]{\includegraphics[width=3.0in]{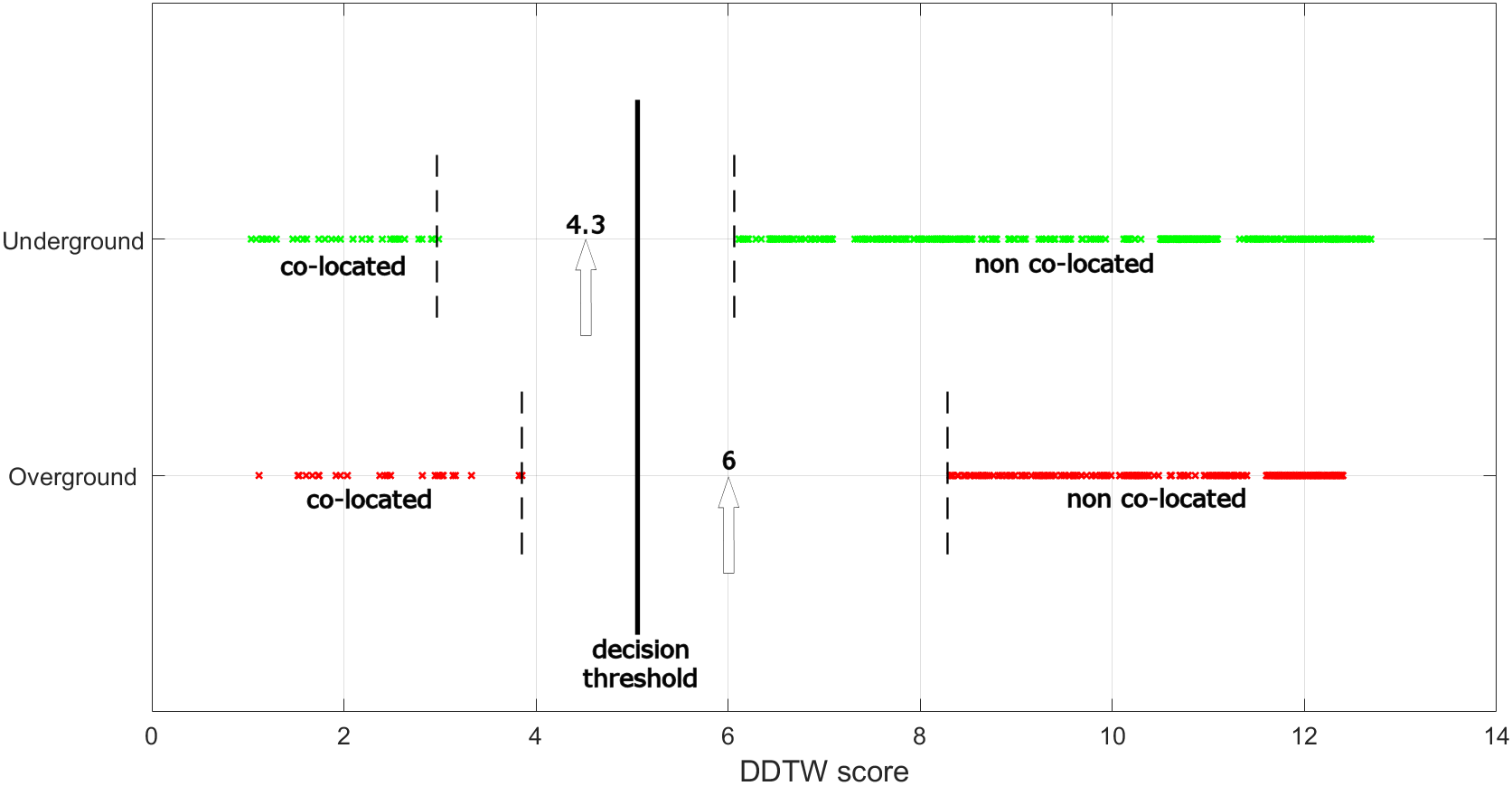}
		\label{ddtwscore}}
	\hfil
	\subfloat[The compress rate heuristic is based on the trajectory length only, hence allowed some false positives.]{\includegraphics[width=3.0in]{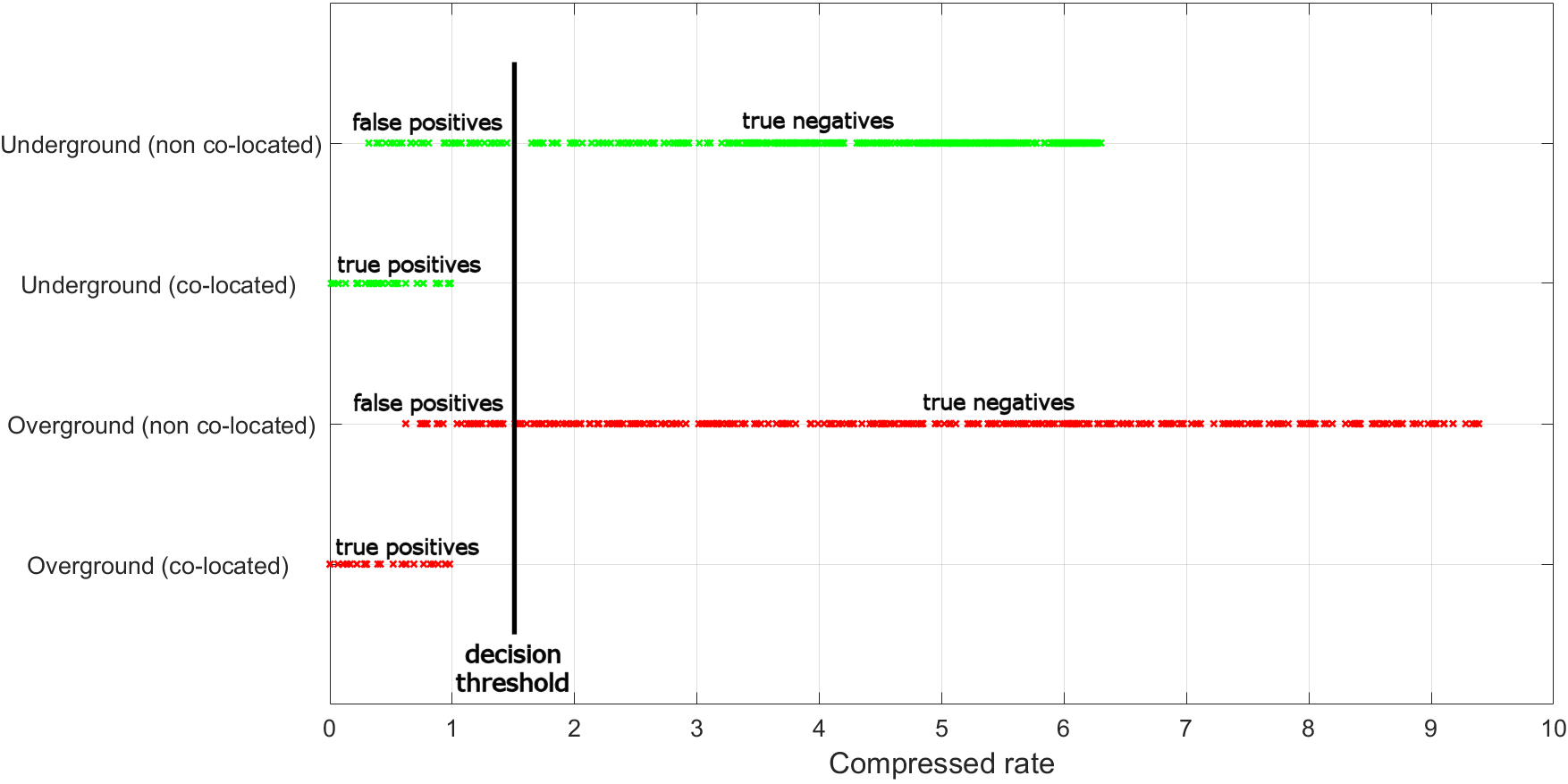}
		\label{compressrate}}	

	\caption{Validating the 676 matching pairs of overground trajectories and 1,156 pairs of underground trajectories. The compress rate heuristic responses much faster than the DDTW score heuristic, albeit allowing some false positives. Thus, we should apply it first to get rid of the majority of the true negatives, then use the DDTW score to get rid of the remaining false positives.}
	\label{decisionthreshold}
\end{figure}

Thus far, we have used visual cues to reinforce the feasibility of using magnetism for co-location detection. The last experiment will inspect the accuracy of our automatic detection algorithm outlined in Section III. There are 26 overground test trajectories and 34 underground ones that connects two consecutive stations for each surveyor. For the sake of testing, we ignore the time-stamp so that our algorithm must only work with the magnetic readings. For each of Alice's trajectories, we compare it to all of Bob's. Our hypothesis is that our algorithm should only accept one of Bob's trajectory - the one that co-locates with Alice's.
Out of a total of 676 pairs of overground trajectories between Alice and Bob, our algorithm correctly identifies all 26 pairs that are indeed co-located. With these co-located pairs, the maximum DDTW score was only 3.8 and the maximum compressed rate was only 1.2. Recalling the heuristics that we defined earlier, these pairs of trajectories satisfied them with wide margins (see Figure~\ref{decisionthreshold}). For the remaining 650 pairs of non-co-located trajectories, our algorithm comfortably rejected them based on just the DDTW distance and the compressed rate criteria. With these pairs, the minimum DDTW score was 8.2 and the maximum compressed rate was 9.4. A similar result was observed with the underground trajectories. Hence, our hypothesis holds for this experiment. It is worth noting that we deliberately ignored the time stamp constraint for this experiment. Realistically, this essential information will help getting rid of many trajectories which start at different times in the real-world.

\subsection{Bus test environment}
\begin{figure}[!t]
	\centering
	\includegraphics[width=3.5in]{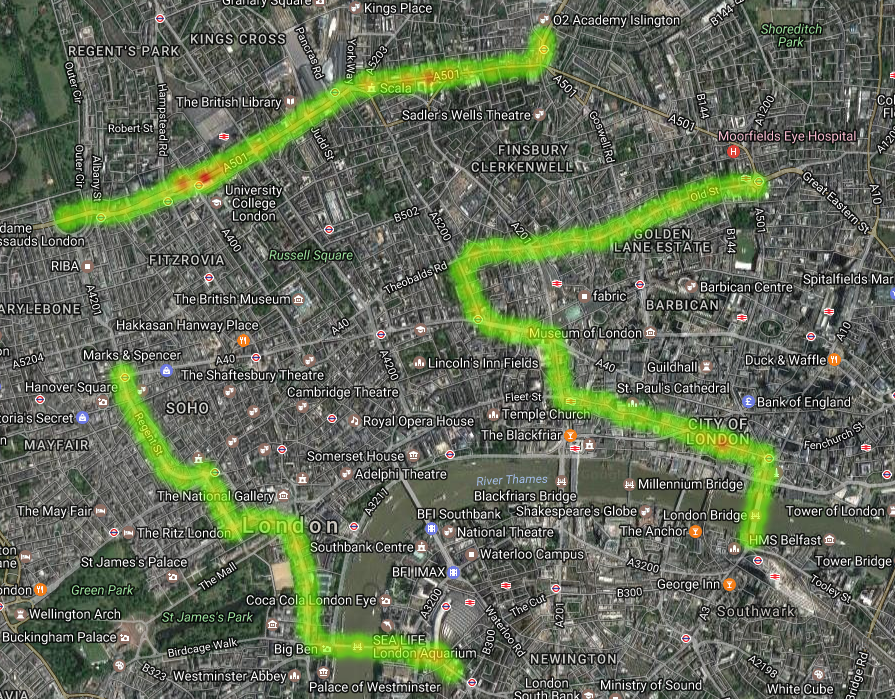}
	\caption{The heatmap of the bus test environment visualised on Google Maps. Regrettably, the magnetic distortion is almost non-existent.}
	\label{busmap}
\end{figure}

Our bus test scenario composes of 3 separate routes, which traverse 22 kilometres of travelling distance in the South-East and Central London (London Bridge - Old Street, Waterloo - Oxford Circus, Regent's Park - Angel), using 4 different buses (see Figure~\ref{busmap}). On top of that, each bus may have an upper deck and a lower deck, which are equivalent to two train carriages. Through-out the experiment, two surveyors sat on different seats on the bus in both decks.

\begin{figure}[!t]
	\centering
	
	\subfloat[London Bridge - Old Street.]{\includegraphics[width=3.0in]{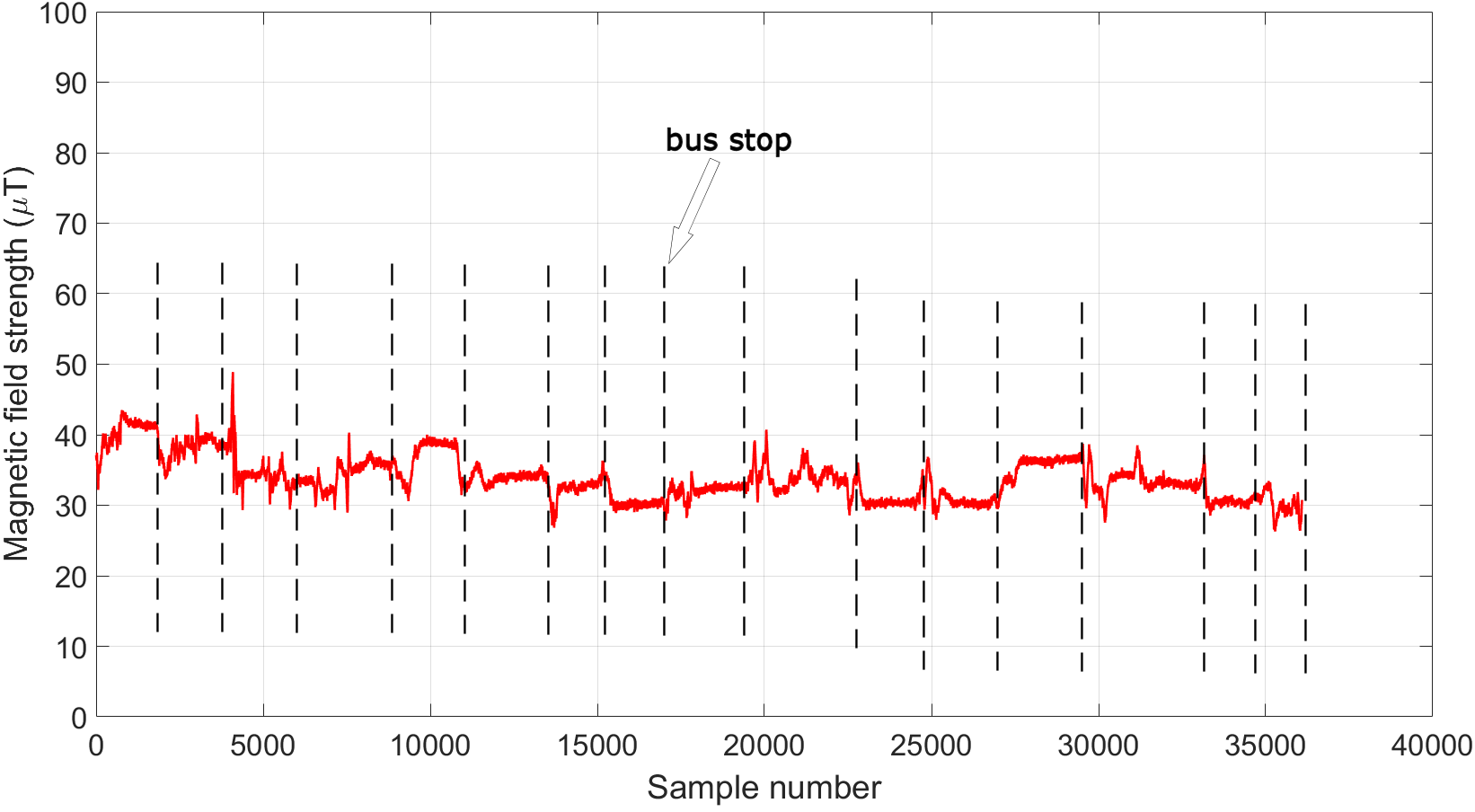}
		\label{bus1}}
	\hfil
	\subfloat[Waterloo - Oxford Circus.]{\includegraphics[width=3.0in]{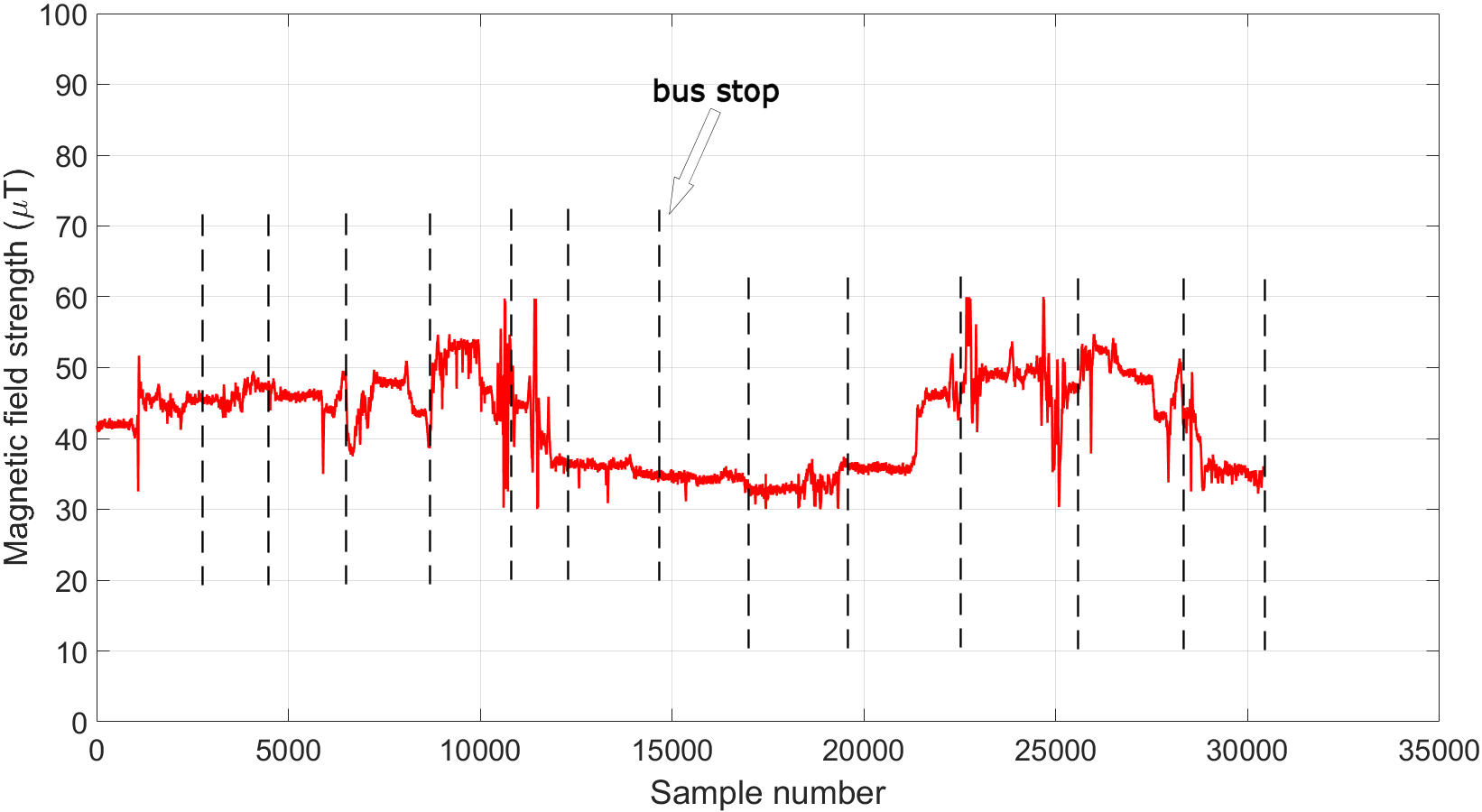}
		\label{bus2}}	
	\hfil
	\subfloat[Regent's Park - Angel.]{\includegraphics[width=3.0in]{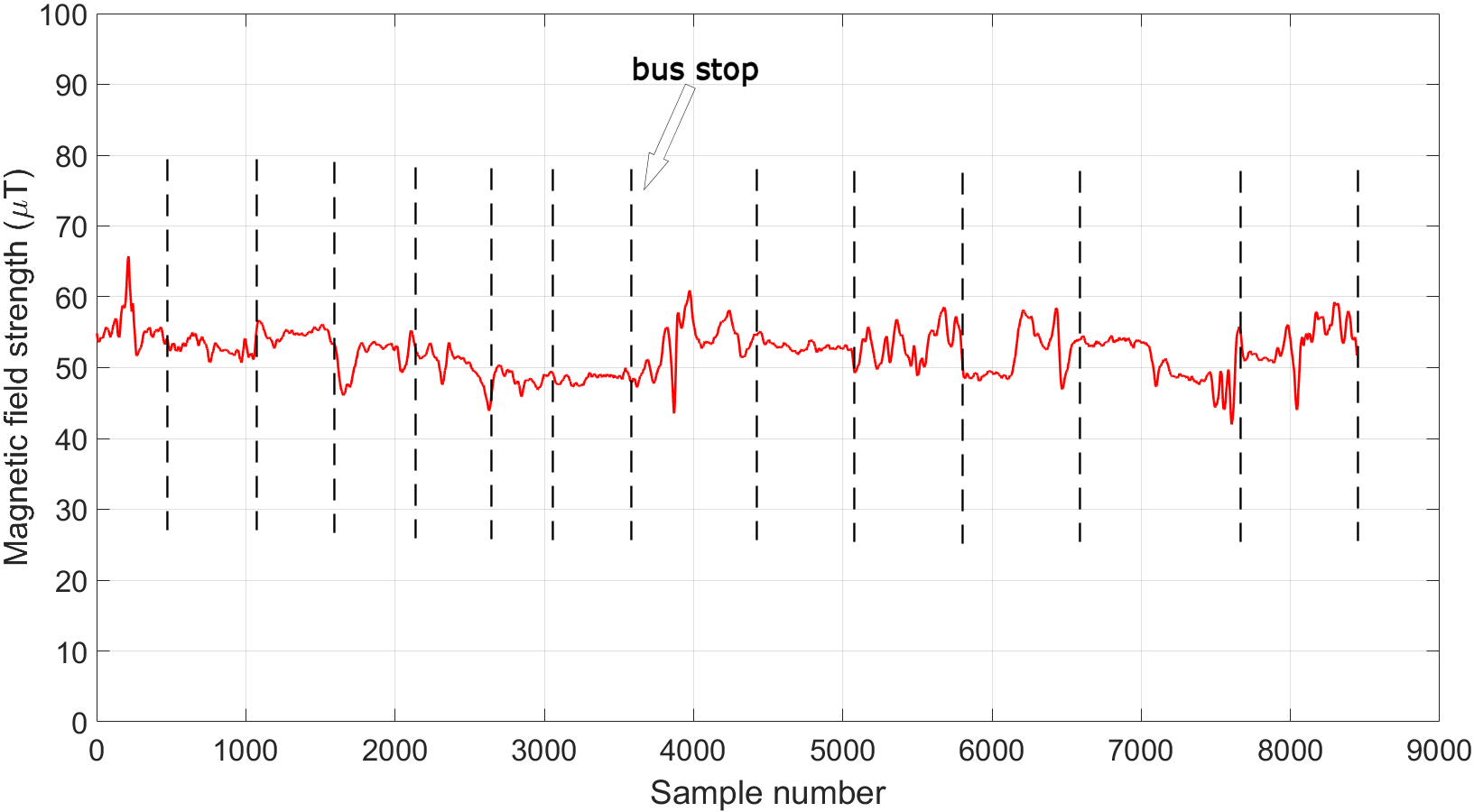}
		\label{bus3}}
	
	\caption{The magnetic field strength between consecutive bus stops. The magnetism variation was considerably less than the previous two train test scenarios. A relatively flat line was observed for multiple trajectories, which denied the chance to co-locate people on the buses.}
	\label{busmagneticvariation}
\end{figure}

Regrettably, a plot of the magnetism from all routes reveals little to no spatial variation. For instance, a 7 minute ride from Lancaster Place stop to Charing Cross stop, passing by 3 different stops had almost zero variation (see Figure~\ref{busmagneticvariation}). The highest magnetic distortion was just 80 $\mu T$ which was observed right in front of Cannon Street station, compared to that of 350 $\mu T$ for the underground test scenario and 210 $\mu T$ for the overground test scenario. 

These results draw up a conclusions that it was not feasible to detect co-location of people on the London buses using magnetism. An empirical explanation is that London buses are hybrid diesel-electric vehicles. They use a diesel engine with electric storage through a lithium ion battery pack. As such, the vehicle itself does not alter the on-board magnetic field much. Additionally, the roads and pavements are a concrete mix of cement and sand which have almost zero impact on magnetism.

\subsection{Summary}
We have presented our empirical experiments to co-locate people on the London public transports. Table~\ref{testbedcompare} summaries the key highlights of our test environments. The clear differences of the underground test environment over the overground one is that the tubes run much faster at almost double the speed. Much more importantly, the level of underground magnetism distortion was much higher than that from the overground trains, which compensates for the short trip length between two consecutive stations. Our approach was very much feasible for the overground trains and the underground tubes, for which the supported railway structures contributed immensely to the high variation of the magnetic field. In contrast, the experiments on the London buses which run on hybrid diesel-electric engine showed little to no magnetic distortion through-out many London routes.
\begin{table}[h]
	\caption{Key highlights of our test environments.}
	\centering 
	\begin{tabular}{| L{2.1cm} | c | c | c |} 
		\hline\hline  
		& \textbf{Overground} & \textbf{Underground} & \textbf{Bus} \\
		& \textbf{train} & \textbf{tube} & \textbf{} \\
		 [0.5ex] 
		\hline 
		Average speed & 30 km/h & 60 km/h & 20 km/h  \\
		Carriage length & 20.4 m & 16.1 m & 11.1 m \\
		Carriage width & 2.8 m & 2.6 m & 2.5 m \\
		Distance coverage & 70 km & 57 km & 22 km \\
		Max coaches & 8 & 7 & 2 \\
		Magnetism variation & Moderate & High & Low \\
		Power & Electricity & Electricity & Diesel-electric \\		
		Shortest trip & 5 minutes & 1 minute & 1 minute \\
		Total stations & 31 & 39 & 42 \\
		\hline 
	\end{tabular}
	\label{testbedcompare}
\end{table}

\section{Related work}
Since the essence of our paper is co-location detection for epidemic tracking, we will only overview other related work in the same area.

Kuk et al detect carriage level co-location of people using just the accelerometer on the smart phone~\cite{kuk2016car}. Their assumption is when the train starts moving, its coaches accelerate differently, which indicates whether two persons are in the same carriage. This is an interesting solution. However, there are two minor impracticalities. Firstly, many people rushes onto the train at the beginning of the trip, and often pro-actively moves to the door before the train reaches its destination. These unexpected movements add a lot of biases to the accelerometer readings, which were not considered in their paper. Secondly, certain trains are pre-programmed so that they accelerate and de-accelerate automatically, which makes it harder to differentiate amongst passengers travelling simultaneously on different trains.

Some of the earliest work in epidemic tracking was from Eiko et al, for which a flu detection system was developed based on GPS and Bluetooth proximity detection~\cite{yoneki2011fluphone}. This type of system actively monitors the user positions in real time, which may be a bit intrusive. Our approach are off-line based monitoring, where the users have completely control to decide whether to upload their personal data for analysis. Additionally, we used low power sensors where Eiko et al used high power sensors. Similarly, Liu et al proposed the same idea using Bluetooth on the smart phones~\cite{liu2014face}, Farrahi et al used Cellular mobile signal~\cite{farrahi2014epidemic}, whereas Nguyen et al used the WiFi signals~\cite{nguyen2015feasibility}.

A complete non-physical epidemic prediction approach was introduced by Coviello et al and Lopes et al~\cite{coviello2016predicting,lopes2009automated}. They relied on the friendship and family ties reported through the social networking databases to predict the spreading rate of a disease. Similarly, Huang et al experiments the flu outbreak using social networking sites in China, using Dynamic Bayesian Network as the underlying algorithm~\cite{huang2013detecting}.

\section{Conclusion and further work}
Verifying if and when two persons are on the same public transport is of paramount importance to contain a disease in an event of epidemic. We have presented an approach to detect co-location of people on the London public transports. The novelty of our work is the use of just low power magnetometer of the smart phone. No GPS, WiFi, Bluetooth or Cellular wireless signals is needed. We have assessed our proposal on the overground trains, the underground tubes and the buses to confirm the feasibility of co-location detection on the trains and the tubes. The buses, on the other hand, did not yield much magnetism variation. To automate the matching process of the user's trajectories, we outlined 4 steps to smooth the raw data, extract the public transport related trajectories, highlight the pair of matched trajectories across different users, and validate the matching pairs. The empirical results displayed a 100\% successful detection ratio on our test environments.

Knowing whether two persons are co-located is not the end story. The longer they stay together, the more chance of being infectious the victim will be. Our next work shall incorporate this information to greatly enhance the usefulness of epidemic tracking. At the end of the day, the users will be happy to engage and contribute to the system if it can be shown to benefit their healthcare.

\bibliographystyle{IEEEtran}
\bibliography{references}
%


\end{document}